\documentclass[a4]{article}
\usepackage{graphicx} 
\usepackage{cite}
\usepackage{amsmath}

\newcommand{\E}{\mathrm{e}}
\newcommand{\D}{\mathrm{d}}
\newcommand{\I}{\mathrm{i}}
\newcommand{\mathfunc}[1]{\mathop{\mathrm{#1}}}
\newcommand{\commut}[2]{\left[ #1 , #2 \right]}

\title{Finding Exponential Product Formulas of Higher Orders}
\author{Naomichi Hatano$^1$ and Masuo Suzuki$^2$\\
{\small $^1$\textit{Institute of Industrial Science, University of Tokyo, Komaba, Meguro, Tokyo 153-8505, Japan}}\\
{\small e-mail: hatano@iis.u-tokyo.ac.jp}\\
{\small $^2$\textit{Department of Applied Physics, Tokyo University of Science, Kagurazaka, Shinjuku, Tokyo 162-8601, Japan}}\\
{\small e-mail: msuzuki@rs.kagu.tus.ac.jp}}
\date{}

\begin{document}

\maketitle

\begin{abstract}
This article is based on a talk presented at a conference ``Quantum Annealing and Other Optimization Methods" held at Kolkata, India on March 2--5, 2005.
It will be published in the proceedings  ``Quantum Annealing and Other Optimization Methods" (Springer, Heidelberg) pp.~39--70.
\end{abstract}

In the present article, we review the progress in the last two decades of the work on the Suzuki-Trotter decomposition\index{Suzuki-Trotter decomposition}, or the exponential product formula\index{exponential product formula}.
The simplest Suzuki-Trotter decomposition, or the well-known Trotter decomposition~\cite{hatano-ref10,hatano-ref20,hatano-ref30,hatano-ref40}\index{Trotter decomposition} is given by
\begin{equation}\label{hatano-eq10}
\E^{x(A+B)}=\E^{xA}\E^{xB}+\mathfunc{O}(x^2),
\end{equation}
where $x$ is a parameter and $A$ and $B$ are arbitrary operators with some commutation relation $\commut{A}{B}\neq0$.
Here the product of the exponential operators on the right-hand side is regarded as an approximate decomposition of the exponential operator on the left-hand side with correction terms of the second order of $x$.
Mathematicians put Eq.~(\ref{hatano-eq10}) in the form
\begin{equation}\label{hatano-eq20}
\E^{xA}\E^{xB}=\E^{x(A+B)+\mathfunc{O}(x^2)}
\end{equation}
and ask what correction terms appear in the exponent of the right-hand side owing to the product in the left-hand side.
They hence refer to it as an exponential product formula.
(The readers should convince themselves by the Taylor expansion that the second-order correction in Eq.~(\ref{hatano-eq10}) is the same as that in Eq.~(\ref{hatano-eq20}).
The higher-order corrections take different forms.)

We here ask how we can generalize the Trotter formula~(\ref{hatano-eq10}) to decompositions with higher-order correction terms.
We concentrate on the form
\begin{equation}\label{hatano-eq30}
\E^{x(A+B)}=\E^{p_1xA}\E^{p_2xB}\E^{p_3xA}\E^{p_4xB}\cdots
\E^{p_MxB}+\mathfunc{O}(x^{m+1}),
\end{equation}
or equivalently
\begin{equation}\label{hatano-eq40}
\E^{p_1xA}\E^{p_2xB}\E^{p_3xA}\E^{p_4xB}\cdots
\E^{p_MxB}=\E^{x(A+B)+\mathfunc{O}(x^{m+1})}.
\end{equation}
We adjust the set of the parameters $\left\{p_1,p_2,\cdots,p_M\right\}$ so that the correction term may be of the order of $x^{m+1}$.
We refer to the right-hand side of Eq.~(\ref{hatano-eq30}) as an $m$th-order approximant in the sense that it is correct up to the $m$th order of $x$.
(See Appendix~\ref{hatano-app05} for another type of the exponential product formula.)

One of the present authors (M.S.) has studied on the higher-order approximant continually~\cite{hatano-ref40,hatano-ref50,hatano-ref60,hatano-ref70,hatano-ref80,hatano-ref90,hatano-ref100,hatano-ref110,hatano-ref120,hatano-ref130,hatano-ref140,hatano-ref150,hatano-ref160,hatano-ref170,hatano-ref180,hatano-ref190,hatano-ref230,hatano-ref200,hatano-ref250,hatano-ref210,hatano-ref220,hatano-ref240,hatano-ref270,hatano-ref275,hatano-ref280,hatano-ref300,hatano-ref290,hatano-ref310,hatano-ref320,hatano-ref330,hatano-ref335,hatano-ref340,hatano-ref350,hatano-ref360}.
The present article mostly reviews his work on the subject.
We first show the importance of the exponential operator in Sect.~\ref{hatano-sec10} and the effectiveness of the exponential product formula in Sect.~\ref{hatano-sec20}.
We demonstrate the effectiveness in examples of the time-evolution operator in quantum dynamics and the symplectic integrator in Hamilton dynamics.
Section~\ref{hatano-sec30} explains a recursive way of constructing higher-order approximants, namely the fractal decomposition.
We present in Sect.~\ref{hatano-sec40} an application of the fractal decomposition to the time-ordered exponential.
We finally review in Sect.~\ref{hatano-sec50} the quantum analysis, an efficient way of computing correction terms of general orders algebraically.
We can use the quantum analysis for the purpose of finding approximants of an arbitrarily high order by solving a set of simultaneous equations where the higher-order correction terms are put to zero.
We demonstrate the prescription in three examples.
We mention in Appendix~\ref{hatano-app05} a type of the exponential product formula different from the form~(\ref{hatano-eq30});
it contains exponentials of commutation relations.
We give in Appendix~\ref{hatano-app10} a short review on the world-line quantum Monte Carlo method with the use of the Trotter approximation~(\ref{hatano-eq10}).

\section{Introduction: Why do we need the exponential product formula?}
\sectionmark{Introduction}
\label{hatano-sec10}

First of all, we discuss as to why we have to treat the exponential operator and why we need an approximant in order to treat the exponential operator.
The exponential operator appears in various fields of physics as a formal solution of the differential equation of the form
\begin{equation}\label{hatano-eq100}
\frac{\partial}{\partial t}f(t)={\cal M}f(t),
\end{equation}
where $f$ is a function or a vector and ${\cal M}$ is an operator or a matrix.
Typical examples are the Schr\"{o}dinger equation\index{Schroedinger equation}
\begin{equation}\label{hatano-eq102}
\I\frac{\partial}{\partial t}\psi(x,t)={\cal H}\psi(x,t)
\end{equation}
(we put $\hbar=1$ here and hereafter),
the Hamilton equation\index{Hamilton equation} 
\begin{equation}
\frac{\D}{\D t}\left(
\begin{array}{c}
\vec{p}(t) \\
\vec{q}(t)
\end{array}
\right)
={\cal H}
\left(
\begin{array}{c}
\vec{p}(t) \\
\vec{q}(t)
\end{array}
\right),
\end{equation}
(see Eq.~(\ref{hatano-eq160}) below) and the diffusion equation with a potential\index{diffusion equation}
\begin{equation}\label{hatano-eq104}
\frac{\D}{\D t}P(x,t)={\cal L}P(x,t).
\end{equation}
A solution of Eq.~(\ref{hatano-eq100}) is given in the form of the Green's function \index{Green's function} as
\begin{equation}\label{hatano-eq110}
f(t)=G(t;0)f(0)=\E^{t{\cal M}}f(0),
\end{equation}
although it is only a formal solution;
obtaining the Green's function $G(t;0)\equiv \E^{t{\cal M}}$ is just as difficult as solving the equation~(\ref{hatano-eq100}) in any other way.
Another important incident of the exponential operator is the partition function\index{partition function} in equilibrium quantum statistical physics:
\begin{equation}\label{hatano-eq120}
Z=\mathfunc{Tr} \E^{-\beta {\cal H}},
\end{equation}
where ${\cal H}$ is a quantum Hamiltonian.

The exponential operator, however, is hard to compute in many interesting cases.
The most straightforward way of computing the exponential operator~$\E^{x{\cal M}}$ is to diagonalize the operator ${\cal M}$.
In quantum many-body problems, however, the basis of the diagonalized representation is often nontrivial, because we are typically interested in the Hamiltonian with two terms or more that are mutually non-commutative;
for example, the Ising model in a transverse field,
\begin{equation}\label{hatano-eq130}
{\cal H}=-\sum_{\langle i,j \rangle}J_{ij}\sigma^z_i\sigma^z_j
-\Gamma\sum_i\sigma^x_i,
\end{equation}
and the Hubbard model,
\begin{equation}\label{hatano-eq140}
{\cal H}=-t\sum_{\sigma=\uparrow,\downarrow}\sum_{\langle i,j \rangle}
\left(c_{i\sigma}^\dag c_{j\sigma}+c_{j\sigma}^\dag c_{i\sigma}\right)
+U\sum_{i}n_{i\uparrow}n_{i\downarrow}.
\end{equation}
In the first example~(\ref{hatano-eq130}), the quantization axis of the first term is the spin $z$ axis, while that of the second term is the spin $x$ axis.
The two terms are therefore mutually non-commutative.
In the second example~(\ref{hatano-eq140}), the first term is diagonalizable in the momentum space,
% as
%\begin{equation}\label{hatano-eq150}
%{\cal K}\equiv -t\sum_{\sigma=\uparrow,\downarrow}\sum_{\langle i,j \rangle}
%\left(c^\dag_{i\sigma}c_{j\sigma}+c^\dag_{j\sigma}c_{i\sigma}\right)
%=\sum_{\vec{k}}\varepsilon\left(\vec{k}\right)
%c_{\vec{k}}^\dag c_{\vec{k}},
%\end{equation}
whereas the second term is diagonalizable in the coordinate space.
In both examples, each term is easily diagonalizable.
Since one quantization axis is different from the other, the diagonalization of the sum of the terms becomes suddenly difficult.

The same situation arises in chaotic Hamilton dynamics.
Consider a classical Hamiltonian
\begin{equation}\label{hatano-eq155}
H(\vec{p},\vec{q})=K(\vec{p})+V(\vec{q}),
\end{equation}
where $K(\vec{p})$ is the kinetic term and $V(\vec{q})$ is the potential term.
The Hamilton equation is expressed in the form
\begin{equation}\label{hatano-eq160}
\frac{\D}{\D t}\left(
\begin{array}{c}
\vec{p}(t) \\
\vec{q}(t)
\end{array}
\right)
=\left(
\begin{array}{r}
-\frac{\D}{\D\vec{q}}V(\vec{q}) \\
\frac{\D}{\D\vec{p}}K(\vec{p})
\end{array}
\right)
\equiv
\left(
\begin{array}{cc}
& -\hat{V}\cdot \\
\hat{K}\cdot &
\end{array}
\right)
\left(
\begin{array}{c}
\vec{p} \\
\vec{q}
\end{array}
\right),
\end{equation}
where the operators $\hat{K}\cdot$ and $\hat{V}\cdot$ are symbolic ones standing for the operations
\begin{equation}\label{hatano-eq170}
\hat{K}\cdot \vec{p}\equiv\frac{\D}{\D\vec{p}}K(\vec{p})
\qquad\mbox{and}\qquad
\hat{V}\cdot \vec{q}\equiv\frac{\D}{\D\vec{q}}V(\vec{q}).
\end{equation}
Although each operation of $\hat{K}\cdot$ and $\hat{V}\cdot$ is simple enough, the ``Hamiltonian" operator
\begin{equation}\label{hatano-eq180}
{\cal H}\equiv
\left(
\begin{array}{cc}
& -\hat{V}\cdot \\
\hat{K}\cdot &
\end{array}
\right)
\end{equation}
is not easily tractable.
This is because the kinetic part and the potential part,
\begin{equation}\label{hatano-eq190}
{\cal K}\equiv
\left(
\begin{array}{cc}
& \phantom{-\hat{V}\cdot}\\
\hat{K}\cdot &
\end{array}
\right)
\qquad\mbox{and}\qquad
{\cal V}\equiv
\left(
\begin{array}{cc}
& -\hat{V}\cdot \\
\phantom{\hat{K}\cdot}&
\end{array}
\right),
\end{equation}
do not commute with each other;
see an example in Sect.~\ref{hatano-subsec20} below.

To summarize this section, we frequently encounter the situation where the exponential operator of each term, $\E^{xA}$ and $\E^{xB}$, is easily obtained and yet the desired exponential operator $\E^{x(A+B)}$ is hard to come.
This is the situation where the Trotter decomposition~(\ref{hatano-eq10}) becomes useful.

\section{Why is the exponential product formula a good approximant?}
\sectionmark{The exponential product formula}
\label{hatano-sec20}

We discussed in the previous section the importance of the exponential operator and the necessity of a way of treating it.
We here discuss a remarkable advantage of the Trotter approximant to the exponential operator.

Let us first confirm that the Trotter approximant~(\ref{hatano-eq10}) is indeed a first-order approximant.
By expanding the both sides of Eq.~(\ref{hatano-eq10}), we have
\begin{eqnarray}\label{hatano-eq200}
\E^{x(A+B)}&=&I+x(A+B)+\frac{1}{2}x^2(A+B)^2+\mathfunc{O}(x^3)
\nonumber\\
&=&I+x(A+B)+\frac{1}{2}x^2\left(A^2+AB+BA+B^2\right)+\mathfunc{O}(x^3),\\
\label{hatano-eq210}
\E^{xA}\E^{xB}&=&
\left(I+xA+\frac{1}{2}x^2A^2+\mathfunc{O}(x^3)\right)
\left(I+xB+\frac{1}{2}x^2B^2+\mathfunc{O}(x^3)\right)
\nonumber\\
&=&
I+x(A+B)+\frac{1}{2}x^2\left(A^2+2AB+B^2\right)+\mathfunc{O}(x^3),
\end{eqnarray}
where $I$ is the identity operator.
The difference between the two comes from the fact that in the approximant~(\ref{hatano-eq210}), the operator $A$ always comes on the left of the operator $B$.
Hence we obtain
\begin{equation}\label{hatano-eq215}
\E^{xA}\E^{xB}=\E^{x(A+B)+\frac{1}{2}x^2\commut{A}{B}+\mathfunc{O}(x^3)}.
\end{equation}

In the actual application of the approximant, we divide the parameter $x$ into $n$ slices in the form
\begin{equation}\label{hatano-eq220}
\left(\E^{\frac{x}{n}A}\E^{\frac{x}{n}B}\right)^n
=\left[
\E^{\frac{x}{n}(A+B)+\frac{1}{2}\left(\frac{x}{n}\right)^2\commut{A}{B}+\mathfunc{O}\left(\left(\frac{x}{n}\right)^3\right)}
\right]^n
=\E^{x(A+B)+\frac{1}{2}\frac{x^2}{n}\commut{A}{B}+\mathfunc{O}\left(\frac{x^3}{n^2}\right)}.
\end{equation}
Thus the correction term vanishes in the limit $n\to\infty$.
We refer to the integer $n$ as the Trotter number.\index{Trotter number}

Now we discuss as to why we should be interested in generalizing the Trotter approximation.
The Trotter approximant~(\ref{hatano-eq10}) and the generalized one~(\ref{hatano-eq30}), in fact, have a remarkable advantage over other approximants such as the frequently used one
\begin{equation}\label{hatano-eq240}
\E^{x(A+B)}=I+x(A+B)+\mathfunc{O}(x^2).
\end{equation}
The approximant of the form~(\ref{hatano-eq30}) conserves an important symmetry of the system in problems of quantum dynamics and Hamilton dynamics.

In problems of quantum dynamics, the exponential operator, or the Green's function $\E^{-\I t{\cal H}}$ is a unitary operator;\index{unitary operator}
hence the norm of the wave function does not change, which corresponds to the charge conservation.
We here emphasize that the exponential product
\begin{equation}\label{hatano-eq230}
%\E^{it{\cal H}/\hbar}\simeq
\E^{-\I tp_1A}\E^{-\I tp_2B}\E^{-\I tp_3A}\cdots
\E^{-\I tp_MB}
\end{equation}
is also a unitary operator.
The perturbational approximant~(\ref{hatano-eq240}), on the other hand, does not conserve the norm of the wave function;
in fact, the norm typically increases monotonically as the time passes as we demonstrate in Sect.~\ref{hatano-subsec19} below.

In problems of Hamilton dynamics, the time evolution of the Hamilton system conserves the volume in the phase space $\left\{\vec{p},\vec{q}\right\}$, which is called the symplecticity\index{symplecticity} in mathematics.
The exponential product formula, in general, also has the symplecticity.

The time evolution of the Hamilton equation~(\ref{hatano-eq160}) is described by the exponential operator
\begin{equation}\label{hatano-eq250}
\left(
\begin{array}{c}
\vec{p}(t) \\
\vec{q}(t)
\end{array}
\right)
=\E^{t{\cal H}}
\left(
\begin{array}{c}
\vec{p}(0) \\
\vec{q}(0)
\end{array}
\right),
\end{equation}
where ${\cal H}$ is the ``Hamiltonian" operator~(\ref{hatano-eq180}).
The Trotter decomposition approximates the time evolution with the operator
\begin{equation}\label{hatano-eq260}
\E^{t{\cal H}}\simeq
\left(\E^{\frac{t}{n}{\cal K}}\E^{\frac{t}{n}{\cal V}}\right)^n
\end{equation}
with ${\cal K}$ and ${\cal V}$ given by Eq.~(\ref{hatano-eq190}).
The operator $\E^{\frac{t}{n}{\cal K}}$ describes the time evolution over the time slice $t/n$ of a Hamilton system with only the kinetic energy $K(p)$.
It thereby conserves the phase-space volume, so does the operator $\E^{\frac{t}{n}{\cal V}}$.
The whole Trotter approximant therefore conserves the phase-space volume.
This holds for any exponential product formula in the form~(\ref{hatano-eq30}) as well.
Hence the exponential product formula, when used in the Hamilton dynamics, is sometimes called a symplectic integrator.\index{symplectic integrator}

In equilibrium quantum statistical physics, the operator $\E^{-\beta{\cal H}}$ does not have a particular symmetry except the symmetries of the Hamiltonian itself.
The above advantage of the exponential product formula is hence lost when applied to numerical calculations of the partition function $Z=\mathfunc{Tr}\E^{-\beta{\cal H}}$.
In fact, in applying the higher-order decomposition~(\ref{hatano-eq30}) to the world-line quantum Monte Carlo simulation, some of the parameters $\left\{p_1,p_2,\cdots,p_M\right\}$ are negative, which causes the negative-sign problem in systems that usually do not have the negative-sign problem~\cite{hatano-ref365}.\index{negative-sign problem}
The negative-sign problem is the problem that the Boltzmann weight of the system to be simulated becomes negative for some configurations.

Thanks to a recent development of the world-line quantum Monte Carlo simulation~\cite{hatano-ref370},\index{quantum Monte Carlo simulation} the higher-order decomposition is not necessary anymore in some cases;
the simulation is carried out in the limit $n\to\infty$ from the very beginning and hence the order of the correction term does not matter in such cases.
See Appendix~\ref{hatano-app10} for a brief review over the recent development.

\subsection{Example: spin precession}
\label{hatano-subsec19}
The fact that the exponential product formula keeps the symmetry of the system is one of its remarkable advantages.
In the present and next subsections, we demonstrate that this indeed affects numerical accuracy strongly.
In the present subsection, we use a simple example of quantum dynamics, namely the spin precession.

Consider the simple Hamiltonian
\begin{equation}\label{hatano-eq5000}
{\cal H}=\sigma_z+\Gamma\sigma_x=\left(
\begin{array}{cc}
1 & \Gamma \\
\Gamma & -1
\end{array}
\right).
\end{equation}
If we start the dynamics from the up-spin state
\begin{equation}\label{hatano-eq5010}
\psi(0)=\left(\begin{array}{c}
1 \\
0
\end{array}\right),
\end{equation}
the spin precesses around the axis of the magnetic field $\vec{H}=(\Gamma,0,1)$ with the period
\begin{equation}\label{hatano-eq5020}
T=\frac{\pi}{\sqrt{1+\Gamma^2}}.
\end{equation}

\begin{figure}
\begin{center}
\includegraphics[width=0.35\textwidth]{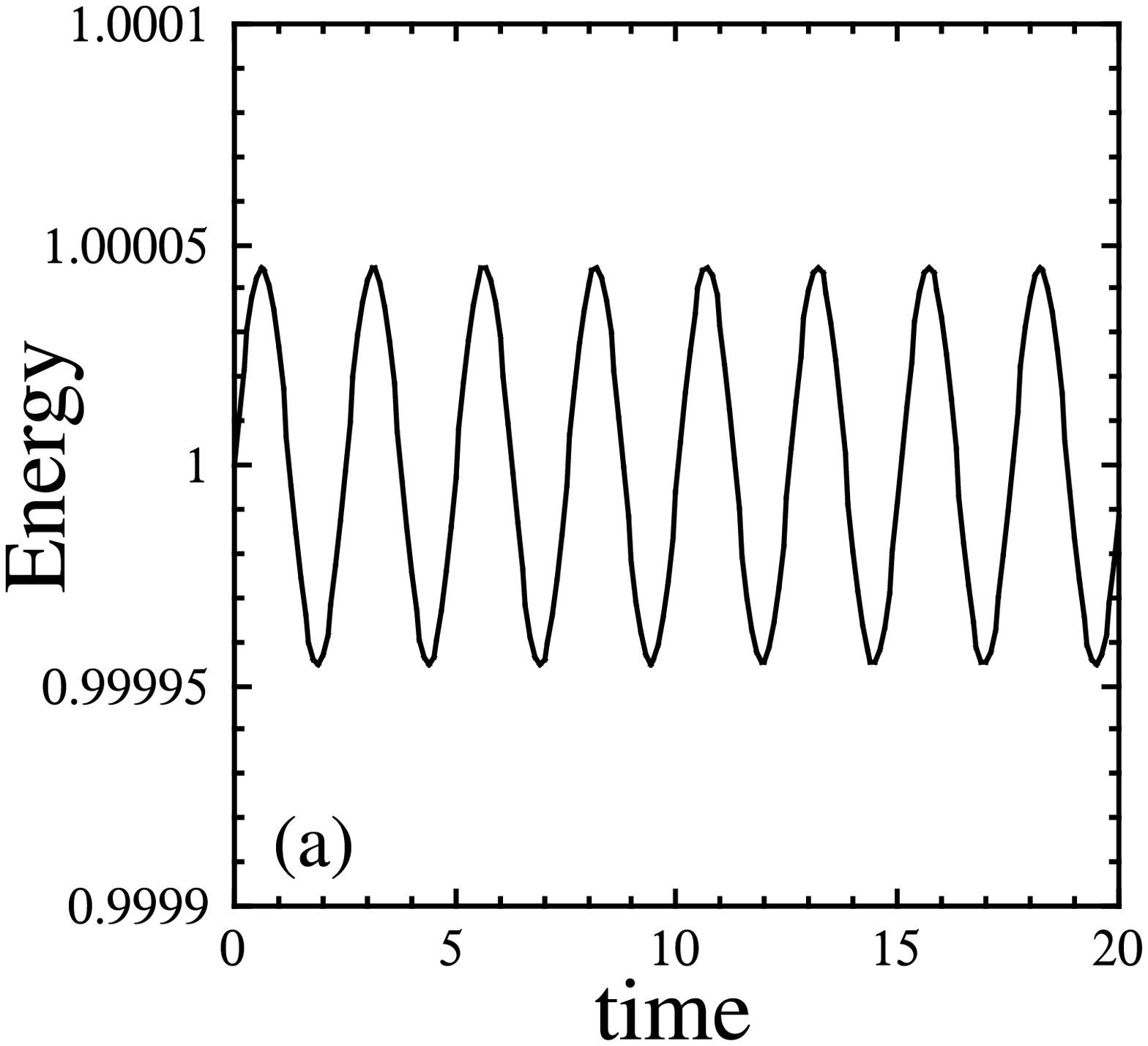}
\hspace{0.1\textwidth}
\includegraphics[width=0.35\textwidth]{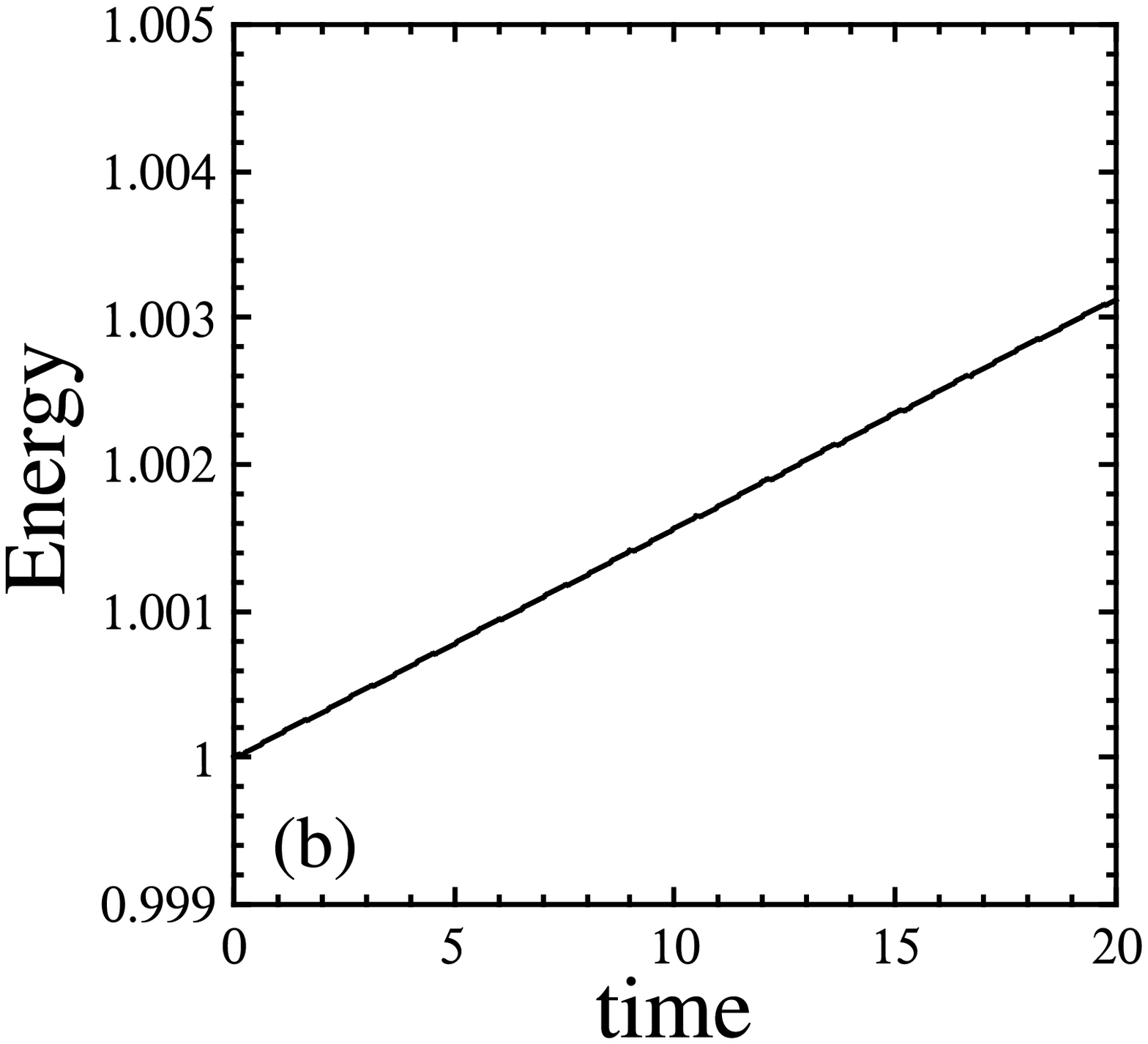}
\end{center}
\caption{The energy deviation due to the approximations given by (a) the Trotter approximant~(\ref{hatano-eq5030}) and (b) the perturbational approximant~(\ref{hatano-eq5040}). In both calculations, we put $\Gamma=3/4$ and $\varDelta t=0.0001$.
The initial state is the one in Eq.~(\ref{hatano-eq5010}) with the energy expectation $\left\langle{\cal H}\right\rangle=1$.}
\label{hatano-fig05}
\end{figure}

Although it is easy to compute the dynamics exactly, we here use the Trotter approximant
\begin{equation}\label{hatano-eq5030}
G(t+\varDelta t;t)\simeq \E^{-\I\varDelta t \sigma_z}\E^{-\I\varDelta t \Gamma\sigma_x}
\end{equation}
and the perturbational approximant\index{perturbational approximant}
\begin{equation}\label{hatano-eq5040}
G(t+\varDelta t;t)\simeq I-\I\varDelta t {\cal H}=I-\I\varDelta t(\sigma_z+\Gamma\sigma_x).
\end{equation}
The exact dynamics should conserve the energy expectation $\left\langle{\cal H}\right\rangle$.
Figure.~\ref{hatano-fig05} shows the energy deviation due to the approximations.
The error in the energy of the Trotter approximation~(\ref{hatano-eq5030}) oscillates periodically and never increases beyond the oscillation amplitude.
The period of the oscillation in Fig.~\ref{hatano-fig05}(a) is equal to that of the spin precession.
We can understand this as follows:
when the spin comes back to the original position after one cycle of the precession, it comes back accurately to the initial state~(\ref{hatano-eq5010}) because of the unitarity of the Trotter approximation, and hence the oscillation.

In contrast, the error in the energy monotonically grows in the case of the perturbational approximant as is shown in Fig.~\ref{hatano-fig05}(b).
This is because the norm of the wave vector increases by the factor
\begin{equation}\label{hatano-eq5050}
\parallel 1-\I\varDelta t{\cal H} \parallel
\simeq 1+\varDelta t \parallel{\cal H}\parallel > 1.
\end{equation}
The remarkable difference between Fig.~\ref{hatano-fig05}(a) and Fig.~\ref{hatano-fig05}(b) thus comes from the fact that the Trotter approximant is unitary.

\subsection{Example: symplectic integrator}
\label{hatano-subsec20}

We next demonstrate the Trotter decomposition~(\ref{hatano-eq260}) in an interesting example of chaotic dynamics.
We again emphasize that keeping the symplecticity of the Hamilton dynamics has an important effect on numerical accuracy.

Let us first notice that the operators in Eq.~(\ref{hatano-eq190}) satisfy
\begin{equation}\label{hatano-eq269}
{\cal K}^2={\cal V}^2=0.
\end{equation}
We therefore have
\begin{eqnarray}\label{hatano-eq270}
\E^{{\cal K}\varDelta t}
\left(
\begin{array}{c}
\vec{p} \\
\vec{q}
\end{array}
\right)
=\left(I+{\cal K}\varDelta t\right)
\left(
\begin{array}{c}
\vec{p} \\
\vec{q}
\end{array}
\right)
=
\left(
\begin{array}{l}
\vec{p} \\
\vec{q}+\varDelta t \frac{\D}{\D\vec{p}}K(\vec{p})
\end{array}
\right),
\\
\label{hatano-eq280}
\E^{{\cal V}\varDelta t}
\left(
\begin{array}{c}
\vec{p} \\
\vec{q}
\end{array}
\right)
=\left(I+{\cal V}\varDelta t\right)
\left(
\begin{array}{c}
\vec{p} \\
\vec{q}
\end{array}
\right)
=
\left(
\begin{array}{l}
\vec{p}-\varDelta t \frac{\D}{\D\vec{q}}V(\vec{q}) \\
\vec{q}
\end{array}
\right).
\end{eqnarray}
Note that applying the two operators in the order $\E^{{\cal K}\varDelta t}\E^{{\cal V}\varDelta t}$ is different from applying them in the order $\E^{{\cal V}\varDelta t}\E^{{\cal K}\varDelta t}$;
in the former, the update of $\vec{q}$ in the application of $\E^{{\cal K}\varDelta t}$ is done under the updated $\vec{p}$, whereas in the latter, it is done under $\vec{p}$ before the update.

\begin{figure}
\begin{center}
\begin{minipage}[t]{0.35\textwidth}
\vspace{0mm}
\includegraphics[width=\textwidth]{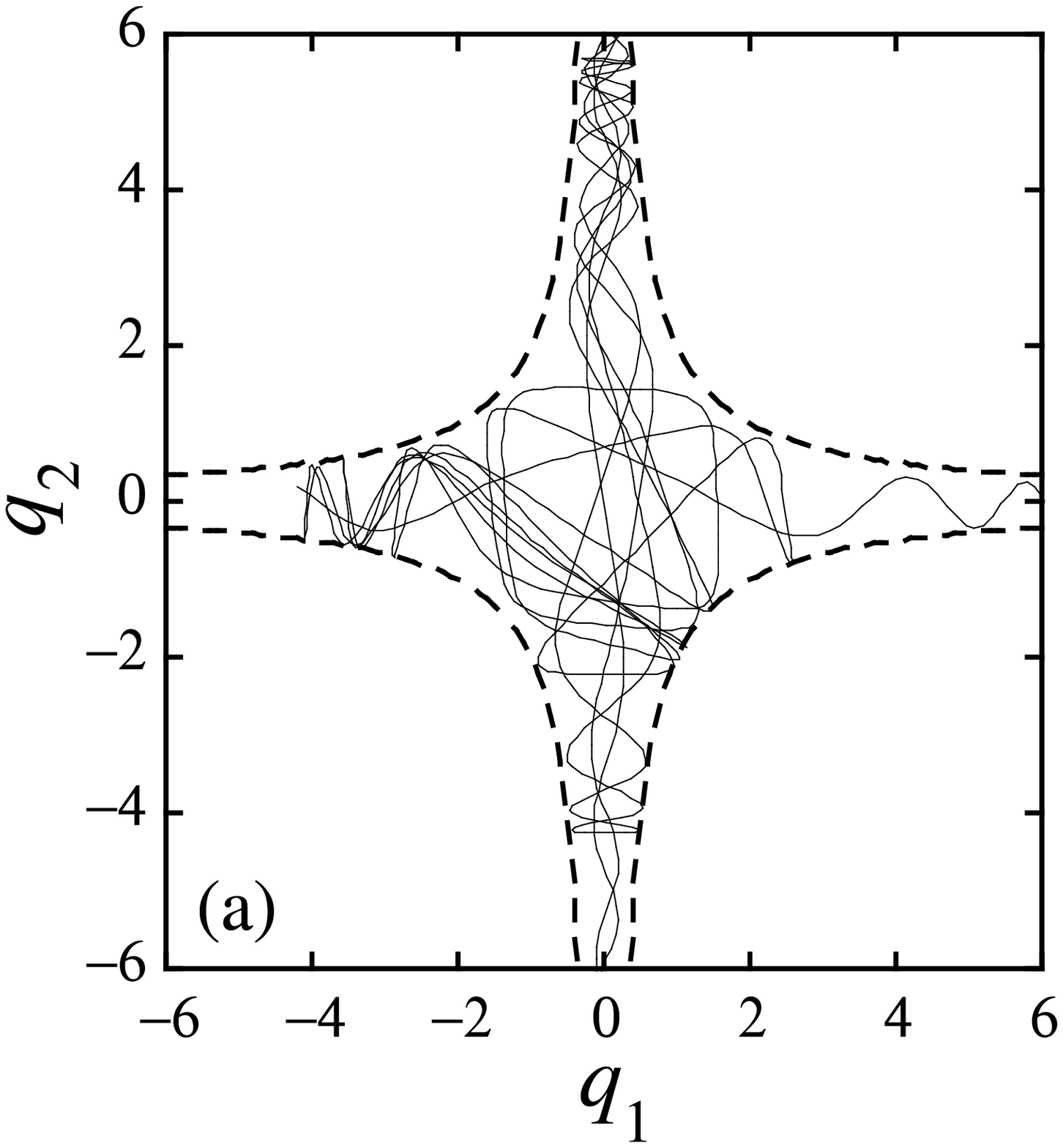}
\end{minipage}
\hspace{0.1\textwidth}
\begin{minipage}[t]{0.4\textwidth}
\vspace{0mm}
\includegraphics[width=\textwidth]{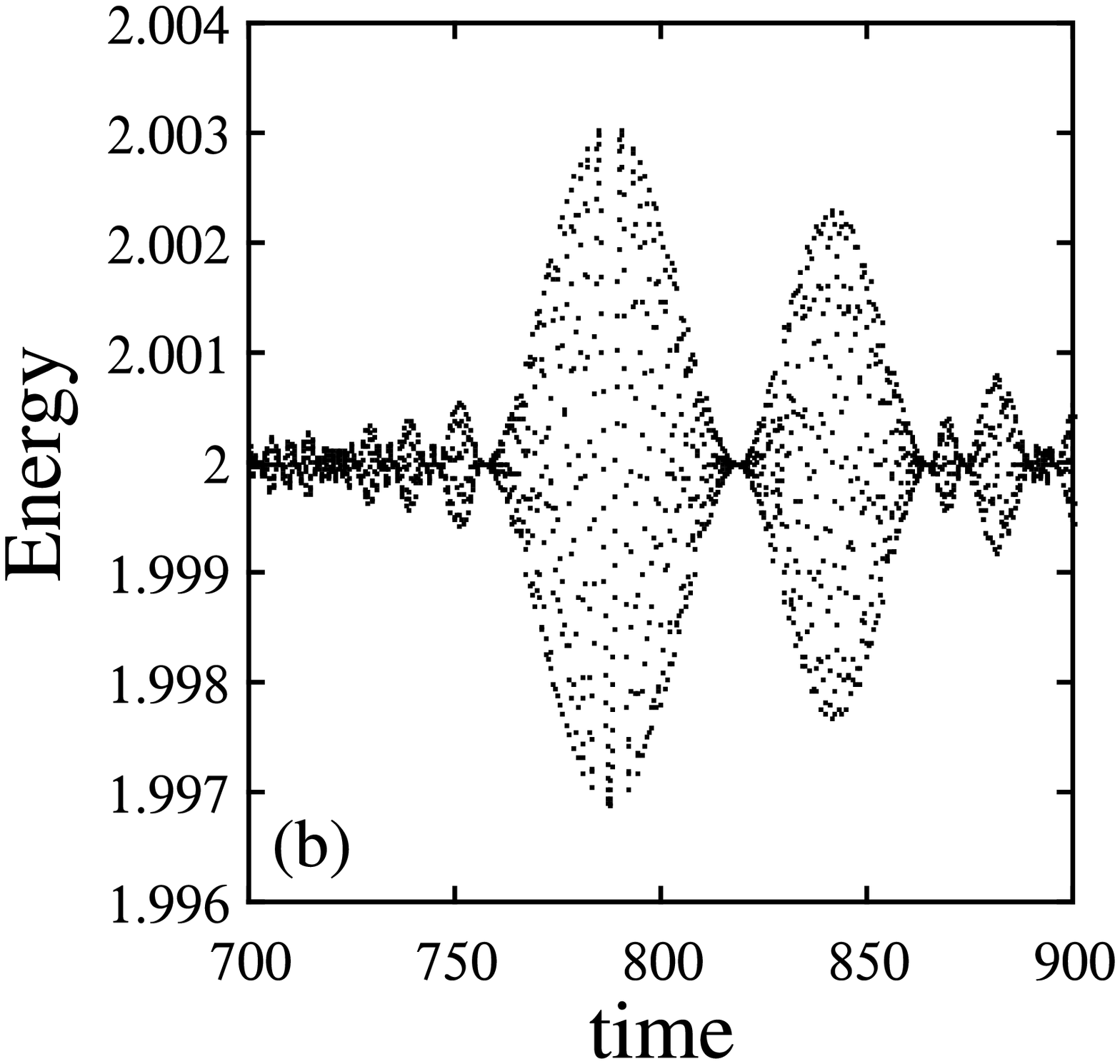}
\end{minipage}
\end{center}
\vspace{\baselineskip}
\begin{center}
\begin{minipage}[t]{0.4\textwidth}
\vspace{0mm}
\caption{Simulations of the system~(\ref{hatano-eq290}).
The initial condition is $p_1=p_2=0$, $q_1=2$ and $q_2=1$ with the energy $E=2$.
The time slice is $\varDelta t=0.0001$.
(a) The movement of the system in the coordinate space $(q_1,q_2)$ for $700\leq t\leq 900$.
The broken curves indicate the hyperbolas $|q_1q_2|=2$.
(b) The energy fluctuation due to the Trotter approximation~(\ref{hatano-eq260}).
We plotted a dot every 1,000 steps.
(c) The energy increase due to the approximant~(\ref{hatano-eq295}).}
\label{hatano-fig10}
\end{minipage}
\hspace{0.05\textwidth}
\begin{minipage}[t]{0.4\textwidth}
\vspace{0mm}
\includegraphics[width=\textwidth]{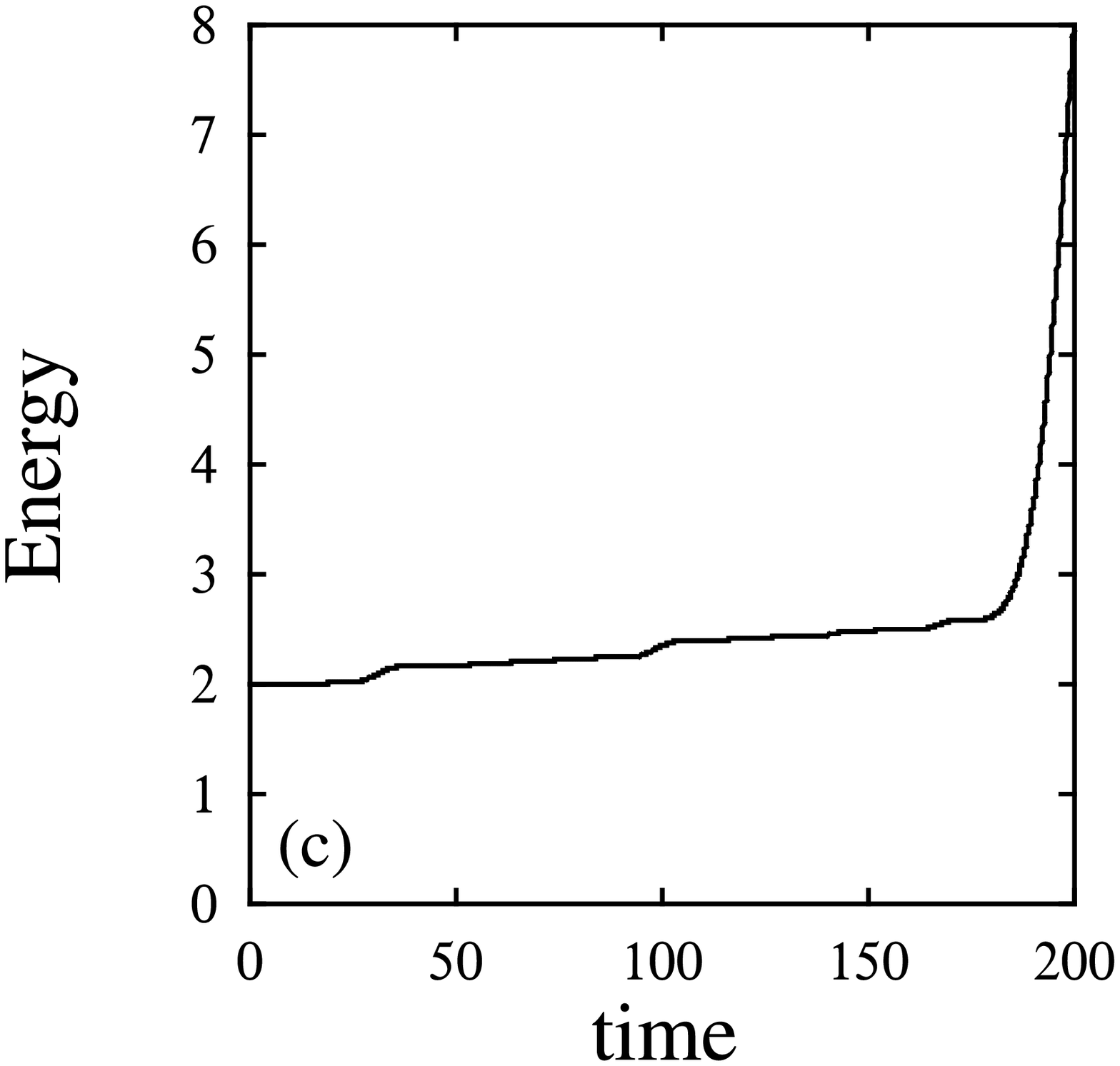}
\end{minipage}
\end{center}
\end{figure}

Umeno and Suzuki~\cite{hatano-ref110,hatano-ref120} demonstrated the use of symplectic integrators\index{symplectic integrator} for chaotic dynamics of the system
\begin{equation}\label{hatano-eq290}
K(\vec{p})=\frac{1}{2}\left(p_1{}^2+p_2{}^2\right)
\qquad\mbox{and}\qquad
V(\vec{q})=\frac{1}{2}q_1{}^2q_2{}^2.
\end{equation}
The equipotential contour is given by $|q_1q_2|=$constant;
hence the system is confined in the area surrounded by four hyperbolas as exemplified in Fig.~\ref{hatano-fig10}(a).
The exact dynamics should conserve the energy.
The Trotter approximation of the dynamics, Eq.~(\ref{hatano-eq260}), gives the energy fluctuation shown in Fig.~\ref{hatano-fig10}(b).
The energy, though deviates from the correct value sometimes, comes back after the deviation.
In fact, the deviation occurs when the system goes into one of the four narrow valleys of the potential;
it is suppressed again and again when the system comes back to the central area.

This is in striking contrast to the update due to the perturbational approximant\index{perturbational approximant}
\begin{equation}\label{hatano-eq295}
\left(
\begin{array}{c}
\vec{p} \\
\vec{q}
\end{array}
\right)
\longrightarrow
\left(I+\varDelta t{\cal H}\right)
\left(
\begin{array}{c}
\vec{p} \\
\vec{q}
\end{array}
\right)
=
\left(
\begin{array}{c}
\vec{p}-\varDelta t\frac{\D}{\D\vec{q}}V(\vec{q}) \\
\vec{q}+\varDelta t\frac{\D}{\D\vec{p}}K(\vec{p})
\end{array}
\right),
\end{equation}
which yields the monotonic energy increase shown in Fig.~\ref{hatano-fig10}(c).
The reason of the difference between the approximants, though less apparent than in the case of the previous subsection, must be keeping the symplecticity, or the conservation of the phase-space volume.

\section{Fractal decomposition}
\label{hatano-sec30}

We emphasized in the previous section the importance of the exponential product formula.
In the present section, we describe a way of constructing higher-order exponential product formulas \textit{recursively}~\cite{hatano-ref50,hatano-ref60,hatano-ref70,hatano-ref80,hatano-ref90,hatano-ref100,hatano-ref110,hatano-ref120,hatano-ref130,hatano-ref140}.\index{fractal decomposition}

The easiest improvement of the Trotter formula~(\ref{hatano-eq20}) is the symmetrization:\index{symmetrized approximant}
\begin{equation}\label{hatano-eq300}
S_2(x)\equiv
\E^{\frac{x}{2}A}\E^{xB}\E^{\frac{x}{2}A}
=\E^{x(A+B)+x^3R_3+x^5R_5+\cdots}.
\end{equation}
The symmetrized approximant has the property
\begin{equation}\label{hatano-eq310}
S_2(x)S_2(-x)=\E^{\frac{x}{2}A}\E^{xB}\E^{\frac{x}{2}A}
\E^{-\frac{x}{2}A}\E^{-xB}\E^{-\frac{x}{2}A}
=I,
\end{equation}
because of which the even-order terms vanish in the exponent of the right-hand side of Eq.~(\ref{hatano-eq300}).
We can thereby promote the approximant~(\ref{hatano-eq300}) to a second-order approximant.

Now we introduce a way of constructing a symmetrized fourth-order approximant from the symmetrized second-order approximant~(\ref{hatano-eq300}).
Consider a product
\begin{eqnarray}\label{hatano-eq305}
S(x)&\equiv& S_2(sx)S_2((1-2s)x)S_2(sx)
\\
\label{hatano-eq306}
&=&\E^{\frac{s}{2}xA}\E^{sxB}\E^{\frac{1-s}{2}xA}\E^{(1-2s)xB}
\E^{\frac{1-s}{2}xA}\E^{sxB}\E^{\frac{s}{2}xA},
\end{eqnarray}
where $s$ is an arbitrary real number for the moment.
The expression~(\ref{hatano-eq300}) is followed by
\begin{eqnarray}\label{hatano-eq320}
S(x)&=&S_2(sx)S_2((1-2s)x)S_2(sx)
\nonumber\\
&=&\E^{sx(A+B)+s^3x^3R_3+\mathfunc{O}(x^5)}
\E^{(1-2s)x(A+B)+(1-2s)^3x^3R_3+\mathfunc{O}(x^5)}
%\nonumber\\
%&&\times
\E^{sx(A+B)+s^3x^3R_3+\mathfunc{O}(x^5)}
\nonumber\\
&=&\E^{x(A+B)+\left[2s^3+(1-2s)^3\right]R_3+\mathfunc{O}(x^5)}.
\end{eqnarray}
(The readers should convince themselves by the Taylor expansion that the third-order correction in the exponent of the last line is just the sum of the third-order corrections in the exponents of the second line.
This is not true for higher-order corrections.)
Note that we arranged the parameters in the form $\{s,1-2s,s\}$ in Eq.~(\ref{hatano-eq305}) so that (i) the first-order term in the exponent of the last line of Eq.~(\ref{hatano-eq320}) should become $x(A+B)$ and (ii) the whole product $S(x)$ should be symmetrized, or should satisfy $S(x)S(-x)=I$.
Because of the second property, the even-order corrections vanish in the exponent of the last line of Eq.~(\ref{hatano-eq320}).
Making the parameter $s$ a solution of the equation
\begin{equation}\label{hatano-eq330}
2s^3+(1-2s)^3=0,
\qquad\mbox{or}\qquad
s=\frac{1}{2-\sqrt[3]{2}}
=1.351207191959657\cdots,
\end{equation}
we promote the product~(\ref{hatano-eq305}) to a fourth-order approximant~\cite{hatano-ref50}.

\begin{figure}
\begin{minipage}[t]{0.25\textwidth}
\vspace{0mm}
\begin{center}
\includegraphics[width=\textwidth]{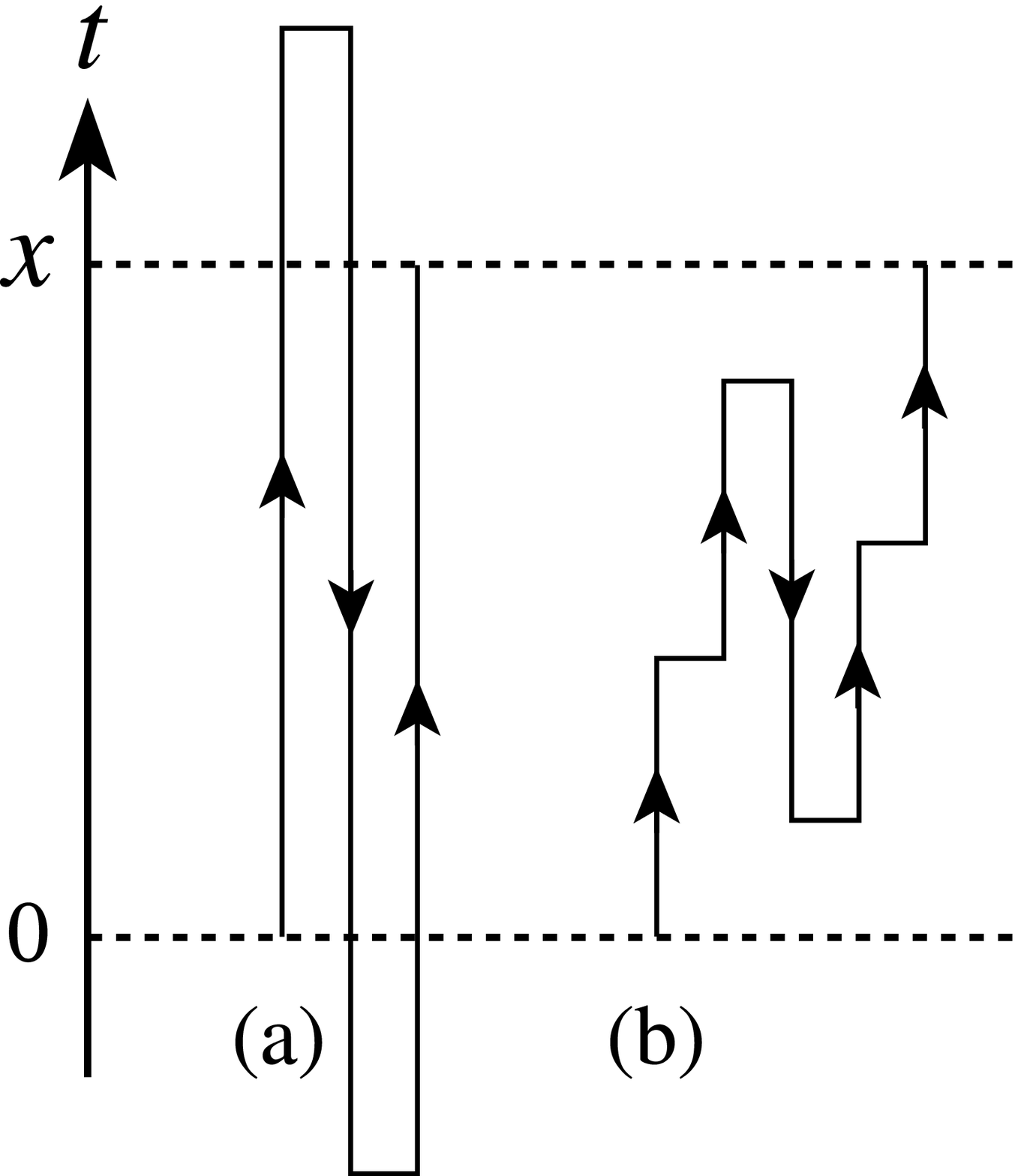}
\end{center}
\end{minipage}
\hfill
\begin{minipage}[t]{0.69\textwidth}
\vspace{0mm}
\begin{center}
\includegraphics[width=\textwidth]{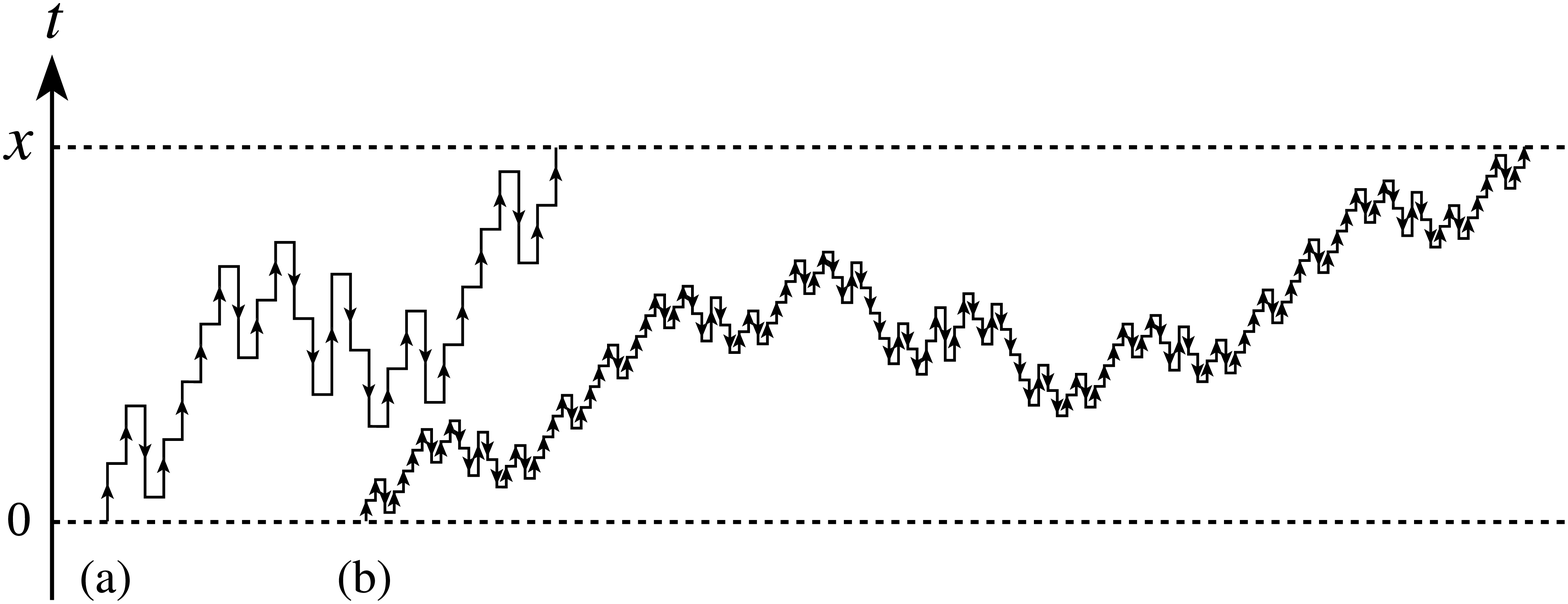}
\end{center}
\end{minipage}

\begin{minipage}[t]{0.3\textwidth}
\vspace{0mm}
\caption{Diagrams that represent the time evolution of (a) the fourth-order approximant~(\ref{hatano-eq305}) and (b) the fourth-order approximant~(\ref{hatano-eq340}).}
\label{hatano-fig20}
\end{minipage}
\hfill
\begin{minipage}[t]{0.6\textwidth}
\vspace{0mm}
\caption{Diagrams that represent the time evolution of (a) the six-order approximant~(\ref{hatano-eq360}) and (b) the eighth-order approximant~(\ref{hatano-eq380}).}
\label{hatano-fig30}
\end{minipage}
\end{figure}

Following the same line of thought, we come up with another fourth-order approximant~\cite{hatano-ref50} in the form
\begin{eqnarray}\label{hatano-eq340}
S_4(x)&\equiv& S_2(s_2x)^2S_2((1-4s_2)x)S_2(s_2x)^2
\\
\label{hatano-eq341}
&=&\E^{\frac{s_2}{2}xA}\E^{s_2xB}\E^{s_2xA}\E^{s_2xB}\E^{\frac{1-3s_2}{2}xA}\E^{(1-4s_2)xB}
\E^{\frac{1-3s_2}{2}xA}
%\nonumber\\
%&&\times
\E^{s_2xB}\E^{s_2xA}\E^{s_2xB}\E^{\frac{s_2}{2}xA},
\end{eqnarray}
where the parameter $s_2$ is a solution of the equation
\begin{equation}\label{hatano-eq350}
4s_2{}^3+(1-4s_2)^3=0,
\qquad\mbox{or}\qquad
s_2=\frac{1}{4-\sqrt[3]{4}}
=0.414490771794375\cdots.
\end{equation}
We can compare the fourth-order approximants~(\ref{hatano-eq305}) and~(\ref{hatano-eq340}) using the following diagram.
Suppose that the exponential operator $\E^{x(A+B)}$ is a time-evolution operator from the time $t=0$ to the time $t=x$.
In the product~(\ref{hatano-eq305}), the term $S_2(sx)$ on the right approximates the time evolution from $t=0$ to $t=sx\simeq1.35x$, the term $S_2((1-2s)x)$ in the middle approximates the time evolution from $t=sx$ to $t=sx+(1-2s)x=(1-s)x\simeq-0.35x$, and the term $S_2(sx)$ on the left approximates the time evolution from $t=(1-s)x$ to $t=(1-s)x+sx=x$.
Let us express this time evolution as in Fig.~\ref{hatano-fig20}(a).
The product~(\ref{hatano-eq340}) is similarly represented as in Fig.~\ref{hatano-fig20}(b).

As is evident, the first product~(\ref{hatano-eq305}) has a part that goes into the ``past," or $t<0$.
This can be problematic in some situations;
in the diffusion\index{diffusion equation} from a delta-peak distribution, for example, there exists no ``past" of the initial delta peak.
The second product~(\ref{hatano-eq340}) does not have the problem and hence is recommended for general use.

Once we know how to construct the fourth-order approximant from the second-order approximant, the rest is quite straightforward~\cite{hatano-ref50}.
Following the construction~(\ref{hatano-eq340}), we construct the sixth-order approximant in the form
\begin{eqnarray}\label{hatano-eq360}
S_6(x)&\equiv&S_4(s_4x)^2S_4((1-4s_4)x)S_4(s_4x)^2
\nonumber\\
&=&\left(S_2(s_4s_2x)^2S_2(s_4(1-4s_2)x)S_2(s_4s_2x)^2\right)^2
\nonumber\\
&&
\times
S_2((1-4s_4)s_2x)^2S_2((1-4s_4)(1-4s_2)x)S_2((1-4s_4)s_2x)^2
\nonumber\\
&&
\times
\left(S_2(s_4s_2x)^2S_2(s_4(1-4s_2)x)S_2(s_4s_2x)^2\right)^2
\end{eqnarray}
with
\begin{equation}\label{hatano-eq370}
4s_4^5+(1-4s_4)^5=0,
\qquad\mbox{or}\qquad
s_4=\frac{1}{4-\sqrt[5]{4}}=0.373065827733272\cdots,
\end{equation}
and further construct the eighth-order approximant in the form
\begin{equation}\label{hatano-eq380}
S_8(x)\equiv S_6(s_6x)^2S_6((1-4s_6)x)S_6(s_6x)^2
\end{equation}
with
\begin{equation}\label{hatano-eq390}
4s_6^7+(1-4s_6)^7=0,
\qquad\mbox{or}\qquad
s_6=\frac{1}{4-\sqrt[7]{4}}=0.359584649349992\cdots.
\end{equation}
These approximants are represented by the diagrams in Fig.~\ref{hatano-fig30}.
%\begin{figure}
%\end{figure}
We can continue this recursive procedure, ending up with the exact time evolution, where the diagram ultimately becomes a fractal object.
This is why the series of the approximants is called the fractal decomposition.\index{fractal decomposition}
It is an interesting thought that the back-and-forth time evolution in a fractal way reproduces the exact time evolution.

\section{Time-ordered exponential}
\label{hatano-sec40}

Before going into another way of constructing higher-order exponential product formulas, let us introduce, as an interlude, an important application of the exponential product formula.
We show how to approximate the time-ordered exponential~\cite{hatano-ref100}.\index{time-ordered exponential}

We have considered until now only the case where the operators $A$ and $B$ do not depend on $x$, or in other words, only the time evolution of a time-independent Hamiltonian.
The fractal decomposition introduced in the previous section needs modification when applied to problems such as the quantum dynamics of a time-dependent Hamiltonian;
in quantum annealing~\cite{hatano-ref380,hatano-ref390,hatano-ref400},\index{quantum annealing} for example, the transverse field $\Gamma$ in the Hamiltonian (\ref{hatano-eq130}) is changed in time.

The time-evolution operator of the quantum Hamiltonian
\begin{equation}\label{hatano-eq400}
{\cal H}(t)=A(t)+B(t)
\end{equation}
is not simply $\E^{-\I{\cal H}t}$ but a time-ordered exponential in the form
\begin{equation}\label{hatano-eq410}
G(t_2;t_1)=\mathfunc{T}\left[\exp\left(-\I\int_{t_1}^{t_2}{\cal H}(s)ds\right)\right].
\end{equation}
%Hereafter, we put $\hbar=1$ for simplicity.
It is quite well-known that
\begin{equation}\label{hatano-eq415}
G_1(t+\varDelta t;t)\equiv\E^{-\I\varDelta tA\left(t+\varDelta t\right)}
\E^{-\I\varDelta tB\left(t+\varDelta t\right)}
\end{equation}
is an approximant of the first order of $\varDelta t$ and
\begin{equation}\label{hatano-eq420}
G_2(t+\varDelta t;t)\equiv \E^{-\frac{\I}{2}\varDelta tA\left(t+\frac{1}{2}\varDelta t\right)}
\E^{-\I\varDelta tB\left(t+\frac{1}{2}\varDelta t\right)}
\E^{-\frac{\I}{2}\varDelta tA\left(t+\frac{1}{2}\varDelta t\right)}
\end{equation}
is an approximant of the second order.
How do we construct higher-order approximants?
We here show that a slight modification of the fractal decomposition gives the answer.

The key is to introduce a shift-time operator~\cite{hatano-ref100}\index{shift-time operator} defined in
\begin{equation}\label{hatano-eq430}
F(t)\E^{-\I\varDelta t{\cal T}}G(t)=F(t+\varDelta t)G(t).
\end{equation}
Note that the operator acts on the function on the left.
The shift-time operator is expressed in the form
\begin{equation}\label{hatano-eq433}
{\cal T}=\I\stackrel{\leftarrow}{\frac{\partial}{\partial t}}
\end{equation}
in the case where $F(t)$ is an analytic function, but the definition~(\ref{hatano-eq430})  does not limit its use to the analytic case.
If we have two shift-time operators, the result is
\begin{eqnarray}\label{hatano-eq435}
F(t)\E^{-\I\varDelta t{\cal T}}G(t)\E^{-\I\varDelta t{\cal T}}H(t)
&=&
F(t+\varDelta t)G(t)\E^{-\I\varDelta t{\cal T}}H(t)
\nonumber\\
&=&
F(t+2\varDelta t)G(t+\varDelta t)H(t).
\end{eqnarray}

With the use of the shift-time operator, the time-ordered exponential~(\ref{hatano-eq410}) is transformed~\cite{hatano-ref100} as
\begin{equation}\label{hatano-eq440}
\mathfunc{T}\left[\exp\left(-\I\int_{t}^{t+\varDelta t}{\cal H}(s)ds\right)\right]
=\E^{-\I\varDelta t\left({\cal H}(t)+{\cal T}\right)}.
\end{equation}
We can prove this by using the Trotter approximation as follows:
\begin{eqnarray}\label{hatano-eq450}
\E^{-\I\varDelta t\left({\cal H}(t)+{\cal T}\right)}
&=&\lim_{n\to\infty}
\left(\E^{-\I\frac{\varDelta t}{n}{\cal H}(t)}
\E^{-\I\frac{\varDelta t}{n}{\cal T}}\right)^n
\nonumber\\
&=&\lim_{n\to\infty}
\E^{-\I\frac{\varDelta t}{n}{\cal H}(t)}
\E^{-\I\frac{\varDelta t}{n}{\cal T}}
\E^{-\I\frac{\varDelta t}{n}{\cal H}(t)}
\E^{-\I\frac{\varDelta t}{n}{\cal T}}\cdots
\E^{-\I\frac{\varDelta t}{n}{\cal H}(t)}
\E^{-\I\frac{\varDelta t}{n}{\cal T}}
\nonumber\\
&=&\lim_{n\to\infty}
\E^{-\I\frac{\varDelta t}{n}{\cal H}\left(t+\varDelta t\right)}
\E^{-\I\frac{\varDelta t}{n}{\cal H}\left(t+\frac{n-1}{n}\varDelta t\right)}
\cdots
%\E^{-\I\frac{\varDelta t}{n}{\cal H}\left(t+\frac{2}{n}\varDelta t\right)}
\E^{-\I\frac{\varDelta t}{n}{\cal H}\left(t+\frac{1}{n}\varDelta t\right)}
\nonumber\\
&=&
\mathfunc{T}\left[\exp\left(-\I\int_{t}^{t+\varDelta t}{\cal H}(s)ds\right)\right].
\end{eqnarray}

Decomposing the Hamiltonian into two parts as in Eq.~(\ref{hatano-eq400}), we have now three parts in the exponent of the time-evolution operator as in
\begin{equation}\label{hatano-eq460}
\mathfunc{T}\left[\exp\left(-\I\int_{t}^{t+\varDelta t}{\cal H}(s)ds\right)\right]
=\E^{-\I\varDelta t\left(A(t)+B(t)+{\cal T}\right)}.
\end{equation}
We then approximate the exponential in the right-hand side of Eq.~(\ref{hatano-eq460}).
The first-order approximant is given by
\begin{eqnarray}\label{hatano-eq462}
G_1(t+\varDelta t;t)
&=&\E^{-\I\varDelta tA(t)}\E^{-\I\varDelta tB(t)}\E^{-\I\varDelta t{\cal T}}
\nonumber\\
&=&\E^{-\I\varDelta tA(t+\varDelta t)}\E^{-\I\varDelta tB(t+\varDelta t)}
\end{eqnarray}
and the second-order approximant is given by
\begin{eqnarray}\label{hatano-eq465}
G_2(t+\varDelta t;t)&=&\E^{-\frac{\I}{2}\varDelta t{\cal T}}
\E^{-\frac{\I}{2}\varDelta tA(t)}
\E^{-\I\varDelta tB(t)}
\E^{-\frac{\I}{2}\varDelta tA(t)}
\E^{-\frac{\I}{2}\varDelta t{\cal T}}
\nonumber\\
&=&
\E^{-\frac{\I}{2}\varDelta tA\left(t+\frac{1}{2}\varDelta t\right)}
\E^{-\I\varDelta tB\left(t+\frac{1}{2}\varDelta t\right)}
\E^{-\frac{\I}{2}\varDelta tA\left(t+\frac{1}{2}\varDelta t\right)}.
\end{eqnarray}
Higher-order approximants are given by the fractal decomposition of the three parts, $A$, $B$, and ${\cal T}$.
The fractal decomposition of three parts is easily obtained by substituting
\begin{equation}\label{hatano-eq470}
S_2(x)\equiv \E^{\frac{x}{2}A}\E^{\frac{x}{2}B}\E^{xC}\E^{\frac{x}{2}B}\E^{\frac{x}{2}A}
=\E^{x(A+B+C)+\mathfunc{O}(x^3)}
\end{equation}
for Eq.~(\ref{hatano-eq300}).
The fourth-order approximant is thereby obtained~\cite{hatano-ref100} as
\begin{eqnarray}\label{hatano-eq480}
%\lefteqn{
G_4(t+\varDelta t;t)
%}
%\nonumber\\
&\equiv&
\left(
\E^{-\frac{\I}{2}s_2\varDelta t{\cal T}}
\E^{-\frac{\I}{2}s_2\varDelta tA(t)}
\E^{-\I s_2\varDelta tB(t)}
\E^{-\frac{\I}{2}s_2\varDelta tA(t)}
\E^{-\frac{\I}{2}s_2\varDelta t{\cal T}}
\right)^2
\nonumber\\
&&\times
\E^{-\frac{\I}{2}(1-4s_2)\varDelta t{\cal T}}
\E^{-\frac{\I}{2}(1-4s_2)\varDelta tA(t)}
%\nonumber\\
%&&\times
\E^{-\I(1-4s_2)\varDelta tB(t)}
\E^{-\frac{\I}{2}(1-4s_2)\varDelta tA(t)}
\E^{-\frac{\I}{2}(1-4s_2)\varDelta t{\cal T}}
\nonumber\\
&&\times
\left(
\E^{-\frac{\I}{2}s_2\varDelta t{\cal T}}
\E^{-\frac{\I}{2}s_2\varDelta tA(t)}
\E^{-\I s_2\varDelta tB(t)}
\E^{-\frac{\I}{2}s_2\varDelta tA(t)}
\E^{-\frac{\I}{2}s_2\varDelta t{\cal T}}
\right)^2
\nonumber\\
&=&
\E^{-\frac{\I}{2}s_2\varDelta tA\left(t+\frac{2-s_2}{2}\varDelta t\right)}
\E^{-\I s_2\varDelta tB\left(t+\frac{2-s_2}{2}\varDelta t\right)}
\E^{-\frac{\I}{2}s_2\varDelta tA\left(t+\frac{2-s_2}{2}\varDelta t\right)}
\nonumber\\
&&\times
\E^{-\frac{\I}{2}s_2\varDelta tA\left(t+\frac{2-3s_2}{2}\varDelta t\right)}
\E^{-\I s_2\varDelta tB\left(t+\frac{2-3s_2}{2}\varDelta t\right)}
\E^{-\frac{\I}{2}s_2\varDelta tA\left(t+\frac{2-3s_2}{2}\varDelta t\right)}
\nonumber\\
&&\times
\E^{-\frac{\I}{2}s_2\varDelta tA\left(t+\frac{1}{2}\varDelta t\right)}
\E^{-\I s_2\varDelta tB\left(t+\frac{1}{2}\varDelta t\right)}
\E^{-\frac{\I}{2}s_2\varDelta tA\left(t+\frac{1}{2}\varDelta t\right)}
\nonumber\\
&&\times
\E^{-\frac{\I}{2}s_2\varDelta tA\left(t+\frac{3s_2}{2}\varDelta t\right)}
\E^{-\I s_2\varDelta tB\left(t+\frac{3s_2}{2}\varDelta t\right)}
\E^{-\frac{\I}{2}s_2\varDelta tA\left(t+\frac{3s_2}{2}\varDelta t\right)}
\nonumber\\
&&\times
\E^{-\frac{\I}{2}s_2\varDelta tA\left(t+\frac{s_2}{2}\varDelta t\right)}
\E^{-\I s_2\varDelta tB\left(t+\frac{s_2}{2}\varDelta t\right)}
\E^{-\frac{\I}{2}s_2\varDelta tA\left(t+\frac{s_2}{2}\varDelta t\right)}
\end{eqnarray}
with the coefficient $s_2$ given by Eq.~(\ref{hatano-eq350}).

\section{Quantum analysis -- Towards the construction of general decompositions --}
\sectionmark{Quantum analysis}
\label{hatano-sec50}

In the last section before the summary, we discuss the calculus of the correction terms.
In the fractal decomposition, we construct higher-order approximants recursively.
Is it possible to construct higher-order approximants \textit{directly}, not recursively?
In fact, Ruth~\cite{hatano-ref375} found (not systematically) a third-order formula\index{Ruth's formula}
\begin{equation}\label{hatano-eq500}
\E^{\frac{7}{24}xA}\E^{\frac{2}{3}xB}\E^{\frac{3}{4}xA}\E^{-\frac{2}{3}xB}\E^{-\frac{1}{24}xA}\E^{xB}=\E^{x(A+B)+\mathfunc{O}(x^4)},
\end{equation}
which would not be found within the framework of the fractal decomposition.

For the purpose of finding higher-order formulas directly, we need to compute the correction terms in the exponent as
\begin{equation}\label{hatano-eq510}
\E^{p_1xA}\E^{p_2xB}\E^{p_3xA}\E^{p_4xB}\cdots
\E^{p_MxB}=\E^{x(A+B)+x^2R_2+x^3R_3+\cdots}.
\end{equation}
This is one of the aims of the quantum analysis\index{quantum analysis} developed by one of the present authors (M.S.)~\cite{hatano-ref300,hatano-ref290,hatano-ref310,hatano-ref320,hatano-ref330,hatano-ref335,hatano-ref340}.
Then we can put the correction terms to zero up to a desired order and solve the set of non-linear simultaneous equations
\begin{equation}\label{hatano-eq520}
R_2=0,
\qquad
R_3=0,
\quad
\cdots,
\quad
R_m=0,
\end{equation}
thereby obtaining the parameters $\{p_i\}$.

\subsection{Operator differential}
\label{hatano-subsec510}

The main feature of the quantum analysis is to introduce operator differential.
In order to motivate the readers, suppose that we can write down an identity
\begin{equation}\label{hatano-eq530}
\frac{\D}{\D x}f(A(x))=\frac{\D f(A)}{\D A}\cdot\frac{\D A(x)}{\D x},
\end{equation}
where $f(A)$ is an operator functional.
The derivative with respect to $x$ on the right-hand side is well-defined;
for example, $\D A(x)/\D x=B+2xC$ for $A(x)=xB+x^2C$.
Now, is it possible to define the differentiation $\D f(A)/\D A$?

Let us discuss as to what should be the definition of the operator differential in order for the identity~(\ref{hatano-eq530}) to hold.
The definition of the $x$ derivative is expressed as
\begin{equation}\label{hatano-eq540}
A(x+h)=A(x)+h\frac{\D A(x)}{\D x}+\mathfunc{O}(h^2).
\end{equation}
The left-hand side of the identity~(\ref{hatano-eq530}) is given by the definition of the derivative as
\begin{equation}\label{hatano-eq550}
%\begin{eqnarray}\label{hatano-eq550}
\frac{\D}{\D x}f(A(x))
=
%&=&
\lim_{h\to0}
\frac{f(A(x+h))-f(A(x))}{h}
%\nonumber\\
%&=&
=
\lim_{h\to0}
\frac{f\left(A(x)+h\frac{\D A(x)}{\D x}\right)-f(A(x))}{h}.
%\end{eqnarray}
\end{equation}
%where
%\begin{equation}\label{hatano-eq555}
%A'(x)\equiv \frac{\D A(x)}{\D x}.
%\end{equation}
The identity~(\ref{hatano-eq530}) suggests that the operator differential $\D f(A)/\D A$ must be a hyperoperator that maps the operator $\D A(x)/\D x$ to the operator given by Eq.~(\ref{hatano-eq550}).

Thus we arrive at the definition of the operator differential within the framework of the quantum analysis~\cite{hatano-ref300}:
if we can express the operator given by
\begin{equation}\label{hatano-eq560}
\D f(A)\equiv\lim_{h\to0}\frac{f(A+h\D A)-f(A)}{h}
\end{equation}
in terms of a hyperoperator mapping from an arbitrary operator $\D A$ as in $\D A \longrightarrow \D f(A)$,
then we refer to the hyperoperator as an operator differential $\D f(A)/\D A$ and denote it in the form
\begin{equation}\label{hatano-eq570}
\D f(A)=\frac{\D f(A)}{\D A}\cdot \D A.
\end{equation}
We stress here that the operator differential $\D f(A)/\D A$ must be expressed in terms of $A$ and the commutation relation of $A$, or the ``inner derivation"
\begin{equation}\label{hatano-eq580}
\delta_A\equiv \commut{A}{\quad},
\end{equation}
but \textit{not} in terms of the arbitrary operator $\D A$.
The convergence of Eq.~(\ref{hatano-eq560}) is in the sense of the norm convergence which is uniform with respect to the arbitrary operator $\D A$.

Let us consider the application of the above in a simple example $f(A)=A^2$.
The definition~(\ref{hatano-eq560}) is followed by
\begin{eqnarray}\label{hatano-eq590}
\D f(A)&=&\lim_{h\to0}\frac{(A+h\D A)^2-A^2}{h}
=\lim_{h\to0}
\frac{hA\,\D A+h\D A\,A+h^2(\D A)^2}{h}
\nonumber\\
&=&
A\,\D A+\D A\,A
=2A\,\D A-(A\,\D A-\D A\,A)
\nonumber\\
&=&\left(2A-\delta_A\right)\D A.
\end{eqnarray}
Thus we have~\cite{hatano-ref300}
\begin{equation}\label{hatano-eq600}
\frac{\D (A^2)}{\D A}=2A-\delta_A.
\end{equation}
%where the inner derivation is defined by Eq.~(\ref{hatano-eq580}).
If $A=xB+x^2C$, we use the result~(\ref{hatano-eq600}) for Eq.~(\ref{hatano-eq530}) and have
\begin{eqnarray}\label{hatano-eq610}
\frac{\D}{\D x}(xB+x^2C)^2&=&\left(2xB+2x^2C-\delta_{xB+x^2C}\right)(B+2xC)
\nonumber\\
&=&(2xB+2x^2C)(B+2xC)-\commut{xB+x^2C}{B+2xC}
\nonumber\\
&=&2xB^2+4x^2BC+2x^2CB+4x^3C^2
%\nonumber\\
%&&
-2x^2(BC-CB)-x^2(CB-BC)
\nonumber\\
&=&2xB^2+3x^2BC+3x^2CB+4x^3C^2,
\end{eqnarray}
which is indeed identical to the result of straightforward algebra.

We cannot see in this simple example any merit of the use of the quantum analysis.
The readers should wait for more complicated examples given later in Sec.~\ref{hatano-subsec530}, where we show that the differential of exponential operators is given in terms of the inner derivation.
The Lie algebra\index{Lie algebra} is defined by commutation relations, or the inner derivation;
it is hence essential to obtain results in terms of the inner derivation, not in terms of naive expansions such as the right-hand side of Eq.~(\ref{hatano-eq610}).

\subsection{Inner derivation}
\label{hatano-subsec520}

We here provide some of the important formulas of the inner derivation~(\ref{hatano-eq580}) as preparation for the next subsection, where we give the differential of exponential operators.

First, we have linearity: for any c-numbers $a$ and $b$, the inner derivation of the operators $A$ and $B$ satisfies
\begin{equation}\label{hatano-eq620}
\delta_{aA+bB}=\commut{aA+bB}{\quad}
=a\commut{A}{\quad}+b\commut{B}{\quad}
=a\delta_A+b\delta_B.
\end{equation}
Any powers of the operator $A$ are commutable with the inner derivation of any powers of the same operator:
\begin{equation}\label{hatano-eq630}
\commut{A^m}{\delta_{A^n}}=0,
\end{equation}
because
\begin{equation}\label{hatano-eq640}
A^m\delta_{A^n}B=A^m\commut{A^n}{B}=\commut{A^n}{A^mB}=\delta_{A^n}A^mB
\end{equation}
for an arbitrary operator $B$ and any integers $m$ and $n$.
We can generalize the identity~(\ref{hatano-eq630}) to the case of any analytic functions of the operator $A$:
\begin{equation}\label{hatano-eq650}
\commut{f(A)}{\delta_{g(A)}}=0,
\end{equation}
where $f(A)$ and $g(A)$ are defined by the Taylor expansion as
\begin{equation}\label{hatano-eq790}
f(A)=\sum_{n=0}^\infty a_nA^n
\qquad\mbox{and}\qquad
g(A)=\sum_{n=0}^\infty b_nA^n.
\end{equation}

Next, we prove the identity~\cite{hatano-ref300}
\begin{equation}\label{hatano-eq660}
\delta_{f(A)g(A)}=f(A)\delta_{g(A)}+g(A)\delta_{f(A)}-\delta_{g(A)}\delta_{f(A)}.
\end{equation}
The proof is as follows: for an arbitrary operator $B$, we have
\begin{eqnarray}\label{hatano-eq670}
%\lefteqn{
f(A)\delta_{g(A)}B+g(A)\delta_{f(A)}B-\delta_{g(A)}\delta_{f(A)}B
%}
%\nonumber\\
&=&
f(A)\commut{g(A)}{B}+g(A)\commut{f(A)}{B}-\commut{g(A)}{\commut{f(A)}{B}}
\nonumber\\
&=&
f(A)\commut{g(A)}{B}+\commut{f(A)}{B}g(A)
=\commut{f(A)g(A)}{B}
\nonumber\\
&=&\delta_{f(A)g(A)}B.
\end{eqnarray}
Note that we can rewrite the identity~(\ref{hatano-eq660}) as
\begin{eqnarray}\label{hatano-eq675}
\delta_{f(A)g(A)}&=&\delta_{g(A)}f(A)+g(A)\delta_{f(A)}-\delta_{g(A)}\delta_{f(A)}
\nonumber\\
&=&\delta_{g(A)}\left(f(A)-\delta_{f(A)}\right)+g(A)\delta_{f(A)}
\end{eqnarray}
because of the identity~(\ref{hatano-eq650}).
In the special case $f(A)=A$, we have
\begin{equation}\label{hatano-eq680}
\delta_{Ag(A)}=\delta_{g(A)}\left(A-\delta_A\right)+g(A)\delta_A.
\end{equation}

With the repeated use of the identity~(\ref{hatano-eq680}), we then prove the identity~\cite{hatano-ref300}
\begin{equation}\label{hatano-eq690}
\delta_{A^n}=A^n-\left(A-\delta_A\right)^n
\end{equation}
for any integer $n$.
This is proved by means of mathematical induction.
The identity~(\ref{hatano-eq690}) indeed holds for $n=1$.
Assume now that
\begin{equation}\label{hatano-eq700}
\delta_{A^{n-1}}=A^{n-1}-\left(A-\delta_A\right)^{n-1}.
\end{equation}
Then the identity~(\ref{hatano-eq680}) yields
\begin{eqnarray}\label{hatano-eq710}
%\lefteqn{
\delta_{A^n}
&=&
\delta_{AA^{n-1}}=\delta_{A^{n-1}}\left(A-\delta_A\right)+A^{n-1}\delta_A
%}
%\nonumber\\
=
\left[A^{n-1}-(A-\delta_A)^{n-1}\right]\left(A-\delta_A\right)+A^{n-1}\delta_A
\nonumber\\
&=&A^n-\left(A-\delta_A\right)^n.
\end{eqnarray}

An interesting and quite well-known identity is
\begin{equation}\label{hatano-eq720}
\E^{xA}B\E^{-xA}=\E^{x\delta_A}B.
\end{equation}
We can prove this by differentiating the left-hand side by $x$.
First, note that
\begin{equation}\label{hatano-eq725}
\frac{\D}{\D x}\E^{xA}B\E^{-xA}
=\E^{xA}AB\E^{-xA}-\E^{xA}BA\E^{-xA}
=\E^{xA}\commut{A}{B}\E^{-xA}.
\end{equation}
We thereby have the following in each order of $x$:
\begin{eqnarray}\label{hatano-eq730}
\left.
\frac{\D}{\D x}\E^{xA}B\E^{-xA}
\right|_{x=0}
&=&
\left.
\E^{xA}\commut{A}{B}\E^{-xA}
\right|_{x=0}
=\delta_A B,
\\
\label{hatano-eq735}
\left.
\frac{\D^2}{\D x^2}\E^{xA}B\E^{-xA}
\right|_{x=0}
&=&
\left.
\E^{xA}\commut{A}{\commut{A}{B}}\E^{-xA}
\right|_{x=0}
={\delta_A}^2 B,
\\
\nonumber
\\
&\cdots,&
\nonumber
\end{eqnarray}
which proves the identity~(\ref{hatano-eq720}).
As a corollary, we obtain the following identity:
\begin{equation}\label{hatano-eq740}
\E^{\delta_A}\E^{\delta_B}=\E^{\delta_\Phi}
\qquad\mbox{if}\qquad
\E^A\E^B=\E^\Phi.
\end{equation}
The proof is straightforward; for an arbitrary operator $C$, we have 
\begin{equation}\label{hatano-eq760}
\E^{\delta_A}\E^{\delta_B}C=\E^{A}\E^{B}C\E^{-B}\E^{-A}=\E^\Phi C \E^{-\Phi}=\E^{\delta_\Phi}C.
\end{equation}

\subsection{Differential of exponential operators}
\label{hatano-subsec530}

We are now in a position to discuss the differential of exponential operators.
We begin with the differential of the power of an operator, $f(A)=A^n$, a generalization of the identity~(\ref{hatano-eq600}).
The result is~\cite{hatano-ref300}
\begin{equation}\label{hatano-eq770}
\frac{\D \left(A^n\right)}{\D A}=\frac{A^n-\left(A-\delta_A\right)^n}{\delta_A}
=\frac{\delta_{A^n}}{\delta_A}.
\end{equation}
An important comment is in order.
The identity~(\ref{hatano-eq770}) does not claim that the inverse of $\delta_A$ is well-defined.
In fact, the inner derivation $\delta_A$ in the denominator is canceled when we expand the numerator of the second expression.
The denominator is well-defined only in such cases.

We use the identity~(\ref{hatano-eq690}) in the derivation of the identity~(\ref{hatano-eq770}).
The definition~(\ref{hatano-eq560}) is followed by
\begin{eqnarray}\label{hatano-eq780}
\D f(A)&=&\lim_{h\to0}\frac{\left(A+h\D A\right)^n-A^n}{h}
=\sum_{j=1}^nA^{j-1}(\D A)A^{n-j}
\nonumber\\
&=&
\left(
nA^{n-1}-\sum_{j=1}^nA^{j-1}\delta_{A^{n-j}}
\right)\D A
%\nonumber\\
%&=&
=
\left\{
nA^{n-1}-\sum_{j=1}^nA^{j-1}
\left[
A^{n-j}-\left(A-\delta_A\right)^{n-j}
\right]
\right\}
\D A
\nonumber\\
&=&
%\left[
\sum_{j=1}^nA^{j-1}\left(A-\delta_A\right)^{n-j}
%\right]
\D A
=\frac{A^n-\left(A-\delta_A\right)^n}{A-\left(A-\delta_A\right)}\D A
\nonumber\\
&=&
\frac{A^n-\left(A-\delta_A\right)^n}{\delta_A}\D A
=\frac{\delta_{A^n}}{\delta_A}\D A.
\end{eqnarray}
Note again that the transformation in the fourth line is well-defined only because the expansion of the numerator cancels the denominator.

We can generalize the identity~(\ref{hatano-eq770}) to any analytic functions defined by the Taylor expansion~(\ref{hatano-eq790}).
The result is
\begin{equation}\label{hatano-eq800}
\frac{\D f(A)}{\D A}=\frac{f(A)-f\left(A-\delta_A\right)}{\delta_A}
=\frac{\delta_{f(A)}}{\delta_A}.
\end{equation}
It is interesting to note that the operator differential or the quantum derivative~\cite{hatano-ref100} is expressed by a difference form of hyperoperators.
As a special case, we arrive at the identity~\cite{hatano-ref300}
\begin{equation}\label{hatano-eq810}
\frac{\D \E^A}{\D A}=\frac{\E^A-\E^{A-\delta_A}}{\delta_A}
=\E^A\frac{1-\E^{-\delta_A}}{\delta_A}.
\end{equation}
%where
%\begin{equation}\label{hatano-eq820}
%\varDelta(A)\equiv
%\frac{\E^{\delta_A}-1}{\delta_A}.
%\end{equation}

\subsection{Example: Baker-Campbell-Hausdorff formula}
\label{hatano-subsec50}

We now use the formula~(\ref{hatano-eq810}) for the derivation of the Baker-Campbell-Hausdorff formula,\index{Baker-Campbell-Hausdorff formula} or the derivation of higher-order terms of the exponent $\Phi(x)$ given in
\begin{equation}\label{hatano-eq830}
\E^{\Phi(x)}=\E^{xA}\E^{xB}.
\end{equation}
The differential of the left-hand side of Eq.~(\ref{hatano-eq830}) gives
\begin{equation}\label{hatano-eq840}
\frac{\D}{\D x}\E^{\Phi(x)}=
\frac{\D \E^{\Phi}}{\D \Phi}\cdot\frac{\D \Phi(x)}{\D x}
=\E^{\Phi(x)}\frac{1-\E^{-\delta_{\Phi(x)}}}{\delta_{\Phi(x)}}\frac{\D \Phi(x)}{\D x}
\end{equation}
owing to Eq.~(\ref{hatano-eq810}), while the differential of the right-hand side of Eq.~(\ref{hatano-eq830}) gives
\begin{equation}\label{hatano-eq850}
%\begin{eqnarray}\label{hatano-eq850}
\frac{\D}{\D x}\E^{xA}\E^{xB}
%&=&
=
\E^{xA}A\E^{xB}+\E^{xA}\E^{xB}B
=\E^{xA}\E^{xB}\left(\E^{-xB}A\E^{xB}+B\right)
%\nonumber\\
%&=&
=
\E^{\Phi(x)}\left(\E^{-x\delta_B}A+B\right),
%\end{eqnarray}
\end{equation}
where we have used the identity~(\ref{hatano-eq720}).
Equating the both sides, we have
\begin{equation}\label{hatano-eq855}
\frac{\D \Phi(x)}{\D x}=
\frac{\delta_{\Phi(x)}}{1-\E^{-\delta_{\Phi(x)}}}\left(\E^{-x\delta_B}A+B\right)
=
\frac{\delta_{\Phi(x)}}{\E^{\delta_{\Phi(x)}}-1}\left(A+\E^{x\delta_A}B\right).
\end{equation}
The second equality is due to the identity~(\ref{hatano-eq740}).

We can expand the right-hand side of Eq.~(\ref{hatano-eq855}) as follows.
Note here that
\begin{equation}\label{hatano-eq860}
\E^{\delta_{\Phi(x)}}=\E^{x\delta_A}\E^{x\delta_B}
\qquad
\mbox{yields}
\qquad
\delta_{\Phi(x)}=\log\left(\E^{x\delta_A}\E^{x\delta_B}\right).
\end{equation}
Thus we transform Eq.~(\ref{hatano-eq855}) as
\begin{equation}\label{hatano-eq870}
%\begin{eqnarray}\label{hatano-eq870}
\frac{\D \Phi(x)}{\D x}
=
%&=&
\frac{\log\left(\E^{x\delta_A}\E^{x\delta_B}\right)}{\E^{x\delta_A}\E^{x\delta_B}-1}\left(A+\E^{x\delta_A}B\right)
=
%\nonumber\\
%&=&
\sum_{k=0}^\infty
\frac{(-1)^k}{k+1}\left(\E^{x\delta_A}\E^{x\delta_B}-1\right)^k\left(A+\E^{x\delta_A}B\right).
%\end{eqnarray}
\end{equation}
We finally arrive~\cite{hatano-ref290} at
\begin{equation}\label{hatano-eq880}
\Phi(x)=\sum_{k=0}^\infty
\frac{(-1)^k}{k+1}
\int_0^x
\left(\E^{t\delta_A}\E^{t\delta_B}-1\right)^k\left(A+\E^{t\delta_A}B\right)dt.
\end{equation}
It is very important to notice here that all the expansion terms are given by commutation relations.
One of the merits of the quantum analysis is to be able to express the expansion in terms of commutation relations.

Let us derive, for example, the term of the third order of $x$, or the second order of $t$ of Eq.~(\ref{hatano-eq880}).
Up to the second order, we have
\begin{eqnarray}\label{hatano-eq890}
\E^{t\delta_A}\E^{t\delta_B}-1&\simeq&
t\left(\delta_A+\delta_B\right)
+\frac{t^2}{2}\left({\delta_A}^2+2\delta_A\delta_B+{\delta_B}^2\right)
\nonumber\\
&=&
t\delta_{A+B}
+\frac{t^2}{2}\left({\delta_{A+B}}^2+\delta_A\delta_B-\delta_B\delta_A\right),
\\
\label{hatano-eq895}
\left(\E^{t\delta_A}\E^{t\delta_B}-1\right)^2&\simeq&
t^2{\delta_{A+B}}^2,
%\nonumber\\
%&=&t^2\left({\delta_A}^2+\delta_A\delta_B+\delta_B\delta_A+{\delta_B}^2\right),
\end{eqnarray}
and hence
\begin{eqnarray}\label{hatano-eq900}
%\lefteqn{
\left(\E^{t\delta_A}\E^{t\delta_B}-1\right)^0\left(A+\E^{t\delta_A}B\right)
&\simeq&
\left(A+B\right)+t\delta_AB+\frac{t^2}{2}{\delta_A}^2B,
%}
\\
\label{hatano-eq902}
%\lefteqn{
\left(\E^{t\delta_A}\E^{t\delta_B}-1\right)^1\left(A+\E^{t\delta_A}B\right)
&\simeq&
t\delta_{A+B}\left(A+B\right)
%}
\nonumber\\
&&
+\frac{t^2}{2}\left({\delta_{A+B}}^2+\delta_A\delta_B-\delta_B\delta_A\right)\left(A+B\right)
+t^2\left(\delta_A+\delta_B\right)\delta_AB
\nonumber\\
&=&\frac{t^2}{2}\left(\delta_A\delta_BA-\delta_B\delta_AB\right)
+t^2\left(\delta_A+\delta_B\right)\delta_AB,
\\
\label{hatano-eq905}
%\lefteqn{
\left(\E^{t\delta_A}\E^{t\delta_B}-1\right)^2\left(A+\E^{t\delta_A}B\right)
%}
%\nonumber\\
%&&
&\simeq&
t^2{\delta_{A+B}}^2\left(A+B\right)=0.
%}
\end{eqnarray}
Summing up the second-order terms with the coefficient $(-1)^k/(k+1)$, we have
\begin{equation}\label{hatano-eq910}
%\lefteqn{
\frac{t^2}{2}{\delta_A}^2B
%}
%\nonumber\\
%&&
-
\frac{t^2}{4}\left(\delta_A\delta_BA-\delta_B\delta_AB\right)
-\frac{t^2}{2}\left(\delta_A+\delta_B\right)\delta_AB
%}
%\nonumber\\
%&=&
=
\frac{t^2}{4}\left({\delta_A}^2B+{\delta_B}^2A\right),
\end{equation}
which we integrate to obtain
\begin{equation}\label{hatano-eq920}
\frac{x^3}{12}\left({\delta_A}^2B+{\delta_B}^2A\right)
=\frac{x^3}{12}\left(
\commut{A}{\commut{A}{B}}
+\commut{\commut{A}{B}}{B}
\right).
\end{equation}

\subsection{Example: Ruth's formula}
\label{hatano-subsec52}

We now extend the above computation to the exponential product
\begin{equation}\label{hatano-eq3000}
\E^{p_1xA}\E^{p_2xB}\E^{p_3xA}\E^{p_4xB}\E^{p_5xA}\E^{p_6xB}=\E^{\Phi(x)}
\end{equation}
and seek Ruth's formula~(\ref{hatano-eq500}) \index{Ruth's formula} as a specific solution of the general formula.
We compute the second-order and third-order correction terms of $\Phi(x)$, defined in
\begin{equation}\label{hatano-eq3010}
\Phi(x)=x(A+B)+x^2R_2+x^3R_3+\mathfunc{O}(x^4),
\end{equation}
and put $R_2=R_3=0$.

The same computation as from Eq.~(\ref{hatano-eq840}) through Eq.~(\ref{hatano-eq880}) produces
\begin{eqnarray}\label{hatano-eq3020}
\Phi(x)&=&\sum_{k=0}^\infty
\frac{(-1)^k}{k+1}
\int_0^x
\left(\E^{p_1t\delta_A}\E^{p_2t\delta_B}\E^{p_3t\delta_A}\E^{p_4t\delta_B}\E^{p_5t\delta_A}\E^{p_6t\delta_B}-1\right)^k
\nonumber\\
&&\phantom{\sum}
\times
\left(p_1A+\E^{p_1t\delta_A}p_2B+\E^{p_1t\delta_A}\E^{p_2t\delta_B}p_3A\cdots\right)dt.
\end{eqnarray}
Note again that all the terms are given by commutation relations.

For the term $k=0$, we have up to the second order of $x$,
\begin{eqnarray}\label{hatano-eq3030}
\lefteqn{
p_1A+\E^{p_1t\delta_A}p_2B+\E^{p_1t\delta_A}\E^{p_2t\delta_B}p_3A\cdots
}
\nonumber\\
&\simeq&
p_1A
+\left(1+tp_1\delta_A+\frac{t^2}{2}{p_1}^2{\delta_A}^2\right)p_2B
\nonumber\\
&&
+\left[1+t\left(p_1\delta_A+p_2\delta_B\right)+\frac{t^2}{2}
\left({p_1}^2{\delta_A}^2
+2p_1p_2\delta_A\delta_B
+{p_2}^2{\delta_B}^2\right)\right]p_3A
%\nonumber\\
%&&
+\cdots
\nonumber\\
&=&
\left(p_1+p_3+p_5\right)A
+\left(p_2+p_4+p_6\right)B
\nonumber\\
&&+t\left[
p_1p_2\delta_AB
+p_2p_3\delta_BA+
\left(p_1+p_3\right)p_4\delta_AB
%\right.
%\nonumber\\
%&&\phantom{+t}\left.
+\left(p_2+p_4\right)p_5\delta_BA
+\left(p_1+p_3+p_5\right)p_6\delta_AB
\right]
\nonumber\\
&&+\frac{t^2}{2}\left[
p_1^2p_2{\delta_A}^2B
+p_2^2p_3{\delta_B}^2A
+2p_1p_2p_3\delta_A\delta_BA
+\left(p_1+p_3\right)^2p_4{\delta_A}^2B
\right.
\nonumber\\
&&\phantom{+x^2}
+2p_2p_3p_4\delta_B\delta_AB
+\left(p_2+p_4\right)^2p_5{\delta_B}^2A
+2\left(p_1p_2+p_1p_4+p_3p_4\right)p_5\delta_A\delta_BA
\nonumber\\
&&\phantom{+x^2}
+\left(p_1+p_3+p_5\right)^2p_6{\delta_A}^2B
%\nonumber\\
%%&&\phantom{+\frac{x^2}{2}}
%%\nonumber\\
%&&\phantom{+x^2}
+2\left(p_2p_3+p_2p_5+p_4p_5\right)p_6\delta_B\delta_AB
\Bigr].
\end{eqnarray}
The zeroth-order term with respect to $t$ appears only here and hence we have the conditions
\begin{equation}\label{hatano-eq3040}
p_1+p_3+p_5=1
\qquad\mbox{and}\qquad
p_2+p_4+p_6=1.
\end{equation}
Using Eq.~(\ref{hatano-eq3040}) and the identity $\delta_BA=-\delta_AB$, we can reduce the right-hand side of Eq.~(\ref{hatano-eq3030}) as
\begin{equation}\label{hatano-eq3050}
%\lefteqn{
A+B+t
(1-2q)
%\left[
%p_1-\left(1-p_1\right)p_2+\left(1-p_2\right)p_3-p_4p_5+p_5p_6
%\right]
\delta_AB
%}
%\nonumber\\
%&&
+\frac{t^2}{2}\left[
%\left[p_1^2p_2-2p_1\left(1-p_1\right)p_2+p_4\left(1-p_5\right)^2-2p_4\left(1-p_5\right)p_5+p_6\right]
(1-q-3r){\delta_A}^2B
%\nonumber\\
%&&%\phantom{+\frac{t^2}{2}}
%\left.
%+\left[p_2^2p_3-2p_2\left(1-p_2\right)p_3+p_5\left(1-p_6\right)^2-2p_5\left(1-p_6\right)p_6\right]
+(q-3s){\delta_B}^2A
\right]
\end{equation}
with
\begin{eqnarray}\label{hatano-eq3052}
q&\equiv&p_2p_3+p_2p_5+p_4p_5,
\\
r&\equiv&p_1p_2p_3+p_1p_2p_5+p_1p_4p_5+p_3p_4p_5,
\\
s&\equiv&p_2p_3p_4+p_2p_3p_6+p_2p_5p_6+p_4p_5p_6.
\end{eqnarray}

For $k=1$, we first have
\begin{equation}\label{hatano-eq3060}
%\begin{eqnarray}\label{hatano-eq3060}
%\lefteqn{
\E^{p_1t\delta_A}\E^{p_2t\delta_B}\E^{p_3t\delta_A}\E^{p_4t\delta_B}\E^{p_5t\delta_A}\E^{p_6t\delta_B}-1
%}
%\nonumber\\
%&\simeq&
\simeq
t\delta_{A+B}
%\nonumber\\
%&&
+\frac{t^2}{2}\left[
{\delta_A}^2+{\delta_B}^2+2(1-q)\delta_A\delta_B
+2q\delta_B\delta_A\right],
%\end{eqnarray}
\end{equation}
where we already used the conditions in Eq.~(\ref{hatano-eq3040}).
Applying Eq.~(\ref{hatano-eq3060}) to Eq.~(\ref{hatano-eq3050}) and dropping higher-order terms, we note that the first-order term vanishes and have
\begin{eqnarray}\label{hatano-eq3070}
\lefteqn{
\left(\E^{p_1t\delta_A}\E^{p_2t\delta_B}\E^{p_3t\delta_A}\E^{p_4t\delta_B}\E^{p_5t\delta_A}\E^{p_6t\delta_B}-1\right)
%}
%\nonumber\\
%&&\times
\left(p_1A+\E^{p_1t\delta_A}p_2B+\E^{p_1t\delta_A}\E^{p_2t\delta_B}p_3A\cdots\right)
}
\nonumber\\
&\simeq&
t^2(1-2q)\delta_{A+B}
%\left[
%p_1-\left(1-p_1\right)p_2+\left(1-p_2\right)p_3-p_4p_5+p_5p_6
%\right]
\delta_AB
%\nonumber\\
%&&
+
\frac{t^2}{2}\left[
{\delta_A}^2+{\delta_B}^2+2(1-q)\delta_A\delta_B
+2q\delta_B\delta_A\right](A+B)
\nonumber\\
&=&
t^2
%\left[
%p_1-\left(1-p_1\right)p_2+\left(1-p_2\right)p_3-p_4p_5+p_5p_6
%\right]
(1-2q)\left({\delta_A}^2B-{\delta_B}^2A\right)
%\nonumber\\
%&&
+\frac{t^2}{2}\left[
{\delta_A}^2B+{\delta_B}^2A-2(1-q){\delta_A}^2B
-2q{\delta_B}^2A
\right]
\nonumber\\
&=&t^2
\left(\frac{1}{2}-q\right)
\left(
%\left[
%\frac{1}{2}-2(1-p_1)p_2-2p_4p_5
%\right]
{\delta_A}^2B
%\right.
%\nonumber\\
%&&\phantom{t^2}\left.
%-\left[
%\frac{1}{2}-p_2p_3-p_5+p_5p_6
%\right]
-{\delta_B}^2A
\right).
\end{eqnarray}

The second-order term of $t$ in the term $k=2$ vanishes just as in Eq.~(\ref{hatano-eq905}).
%because $\delta_{A+B}(A+B)=0$.
Thus we arrive at
\begin{equation}\label{hatano-eq3080}
%\begin{eqnarray}\label{hatano-eq3080}
\Phi(x)
%&=&
=
x(A+B)+\frac{x^2}{2}
%\left[
%p_1-\left(1-p_1\right)p_2+\left(1-p_2\right)p_3-p_4p_5+p_5p_6
%\right]
(1-2q)\delta_AB
%\nonumber\\
%&&
+\frac{x^3}{3!}\left[
%\left[1+2p_1p_2+3p_1^2p_2-4p_2+p_4-8p_4p_5+3p_4{p_5}^2+p_6\right]
\left(\frac{1}{2}-3r\right){\delta_A}^2B
%\right.
%\nonumber\\
%&&%\phantom{+\frac{t^2}{2}}
%\left.
+
%\left[1-4p_2p_3+3p_2^2p_3-p_5-2p_5p_6+3p_5{p_6}^2\right]
\left(\frac{1}{2}-3s\right){\delta_B}^2A
\right]+\mathfunc{O}(x^4).
%\end{eqnarray}
\end{equation}
Putting the second-order and third-order terms to zero, we have a set of simultaneous equations of the parameters as
\begin{eqnarray}\label{hatano-eq3090}
p_1+p_3+p_5&=&1,
\\
\label{hatano-eq3091}
p_2+p_4+p_6&=&1,
\\
\label{hatano-eq3092}
2q=
2\left(
p_2p_3+p_2p_5+p_4p_5
\right)
%p_1-\left(1-p_1\right)p_2+\left(1-p_2\right)p_3-p_4p_5+p_5p_6
&=&1,
\\
\label{hatano-eq3093}
%1+2p_1p_2+3p_1^2p_2-4p_2+p_4-8p_4p_5+3p_4{p_5}^2+p_6
%6\left(
%p_1p_2p_3+p_1p_2p_5+p_1p_4p_5+p_3p_4p_5
%\right)
6r
\stackrel{(\ref{hatano-eq3092})}{=}3\left(p_1+2p_3p_4p_5\right)
&=&1,
\\
\label{hatano-eq3094}
%1-4p_2p_3+3p_2^2p_3-p_5-2p_5p_6+3p_5{p_6}^2
%6\left(
%p_2p_3p_4+p_2p_3p_6+p_2p_5p_6+p_4p_5p_6
%\right)
6s
\stackrel{(\ref{hatano-eq3092})}{=}3\left(2p_2p_3p_4+p_6\right)
&=&1.
\end{eqnarray}
We can confirm that Ruth's formula~(\ref{hatano-eq500}), or
\begin{equation}\label{hatano-eq3100}
p_1=\frac{7}{24},
\;\;
p_2=\frac{2}{3},
\;\;
p_3=\frac{3}{4},
\;\;
p_4=-\frac{2}{3},
\;\;
p_5=-\frac{1}{24},
\quad\mbox{and}\quad
p_6=1
\end{equation}
is indeed a solution of the above set of simultaneous equations.
With six variables for five equations, the solution is in fact a continuous line;
Ruth's solution~(\ref{hatano-eq3100}) is just a point on the line.
By adjusting the last variable $p_6$, we have the continuous solution shown in Fig.~\ref{hatano-fig50}.
(We can solve the set of equations with five parameters by putting $p_6=0$, but the solution is complex.)

\subsection{Example: perturbational composition}
\label{hatano-subsec54}

We finally present an interesting exercise, motivated by the ``perturbational composition"~\cite{hatano-ref378}.\index{perturbational composition}
Suppose that we apply a weak transverse field to an Ising spin.
We ask what is the correction term in the exponent of the right-hand side of
\begin{equation}\label{hatano-eq4000}
\E^{\frac{x}{2}\gamma\sigma_x}\E^{x\sigma_z}
\E^{\frac{x}{2}\gamma\sigma_x}
=\E^{\Phi(x,\gamma)}
=\E^{x\left(\sigma_z+\gamma C_1(x)+\mathfunc{O}(\gamma^2)\right)}.
\end{equation}
Notice that we expand the exponent with respect to the perturbation parameter $\gamma$, not with respect to $x$ as in the preceding sections.
The first-order perturbation term $C_1(x)$ in turn contains higher orders of $x$.
We could explicitly compute the $2\times2$ matrices on both sides of Eq.~(\ref{hatano-eq4000}), expand them with respect to $\gamma$ and compare them term by term, but the quantum analysis provides a more elegant way of computation.

We differentiate the both sides of Eq.~(\ref{hatano-eq4000}) with respect to $\gamma$:
\begin{eqnarray}\label{hatano-eq4010}
\frac{\D}{\D \gamma}\E^{x\left(\sigma_z+\gamma C_1(x)+\mathfunc{O}(\gamma^2)\right)}
&=&
\frac{\D \E^\Phi}{\D \Phi}\cdot\frac{\partial\Phi(x,\gamma)}{\partial\gamma}
=\E^\Phi\frac{1-\E^{-\delta_{\Phi}}}{\delta_\Phi}\frac{\partial\Phi}{\partial\gamma},
\\
\label{hatano-eq4020}
\frac{\D}{\D \gamma}
\E^{\frac{x}{2}\gamma\sigma_x}
\E^{x\sigma_z}
\E^{\frac{x}{2}\gamma\sigma_x}
&=&\frac{x}{2}\left(
\sigma_x
\E^{\frac{x}{2}\gamma\sigma_x}
\E^{x\sigma_z}
\E^{\frac{x}{2}\gamma\sigma_x}
+
\E^{\frac{x}{2}\gamma\sigma_x}
\E^{x\sigma_z}
\E^{\frac{x}{2}\gamma\sigma_x}
\sigma_x
\right)
\nonumber\\
&=&\frac{x}{2}\E^\Phi
\left(\E^{-\Phi}\sigma_x \E^\Phi+\sigma_x\right)
%\nonumber\\
%&=&
=
\frac{x}{2}\E^\Phi\left(\E^{-\delta_\Phi}+1\right)\sigma_x.
\end{eqnarray}
Equating the both sides, we have
\begin{eqnarray}\label{hatano-eq4030}
\frac{\partial\Phi}{\partial\gamma}
%&=&
=
xC_1(x)+\mathfunc{O}(\gamma)
%\nonumber\\
&=&\frac{x}{2}\frac{\delta_\Phi}{1-\E^{-\delta_\Phi}}
\left(1+\E^{-\delta_\Phi}\right)\sigma_x
\nonumber\\
&=&\frac{x}{2}\left(x\delta_{\sigma_z}+\mathfunc{O}(\gamma)\right)
\frac{1+\E^{-\delta_\Phi}}{1-\E^{-\delta_\Phi}}\sigma_x.
\end{eqnarray}
Putting $\gamma=0$, we have
\begin{equation}\label{hatano-eq4040}
C_1(x)=\frac{1}{2}x\delta_{\sigma_z}
\frac{1+\exp\left(-x\delta_{\sigma_z}\right)}{1-\exp\left(-x\delta_{\sigma_z}\right)}
\sigma_x
=\frac{1}{2}\sum_{n=0}^\infty a_nx^n{\delta_{\sigma_z}}^n\sigma_x
\end{equation}
owing to the fact $\delta_\Phi=x\delta_{\sigma_z}+\mathfunc{O}(\gamma)$, where we have made the Taylor expansion
\begin{equation}\label{hatano-eq4050}
x\frac{1+\E^{-x}}{1-\E^{-x}}=\sum_{n=0}^\infty a_nx^n
\end{equation}
with $a_0=1$.
We also note that the function~(\ref{hatano-eq4050}) is even with respect to $x$ and hence $a_{n}=0$ for odd integers $n$.
\begin{figure}
\begin{center}
\begin{minipage}[t]{0.35\textwidth}
\vspace{0mm}
\includegraphics[width=\textwidth]{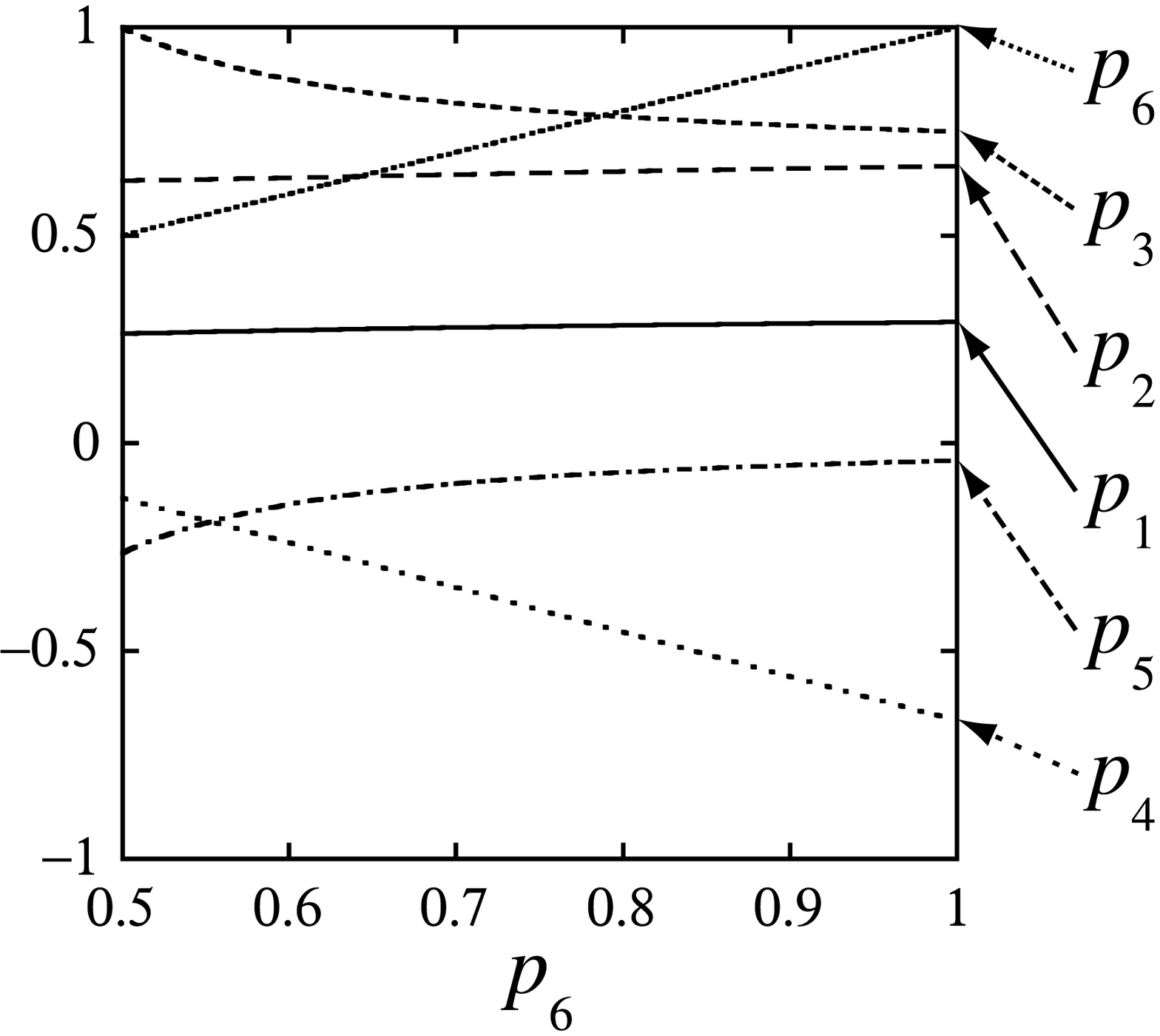}
\caption{The solution line of the set of simultaneous equations~(\ref{hatano-eq3090})--(\ref{hatano-eq3094}).}
\label{hatano-fig50}
\end{minipage}
\hspace{0.1\textwidth}
\begin{minipage}[t]{0.35\textwidth}
\vspace{0mm}
\includegraphics[width=\textwidth]{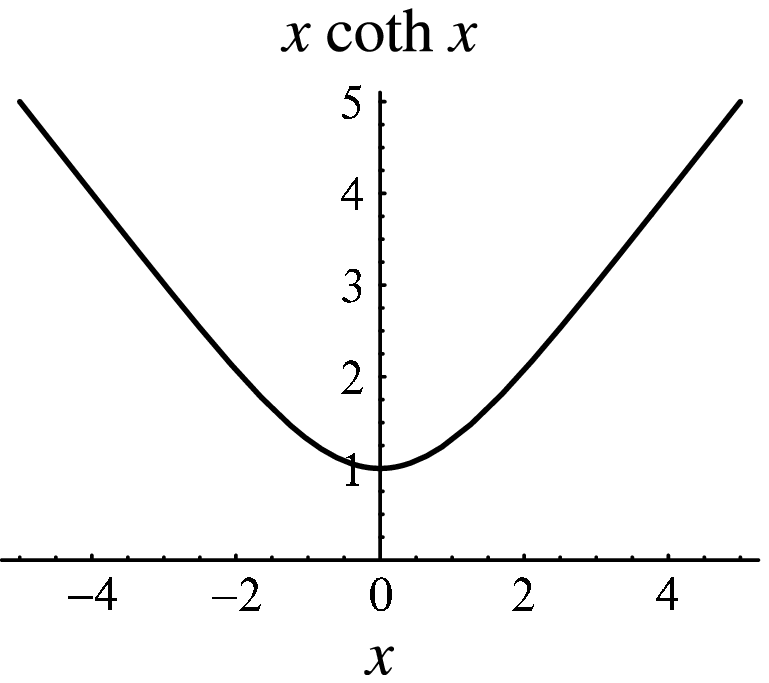}
\caption{The coefficient of the first-order perturbation of Eq.~(\ref{hatano-eq4090}).}
\label{hatano-fig40}
\end{minipage}
\end{center}
\end{figure}

The right-hand side of Eq.~(\ref{hatano-eq4040}) is explicitly calculated as follows.
In each order, we have
\begin{eqnarray}\label{hatano-eq4060}
\delta_{\sigma_z}\sigma_x
&=&
\commut{\sigma_z}{\sigma_x}=2\I\sigma_y,
\\
\label{hatano-eq4062}
{\delta_{\sigma_z}}^2\sigma_x
&=&2\I\commut{\sigma_z}{\sigma_y}=4\sigma_x,
\\
\label{hatano-^eq4064}
{\delta_{\sigma_z}}^3\sigma_x
&=&
4\commut{\sigma_z}{\sigma_x}=8\I\sigma_y,
\\
&&\cdots,
\nonumber
\end{eqnarray}
or in general,
\begin{equation}\label{hatano-eq4070}
{\delta_{\sigma_z}}^n\sigma_x=\left\{
\begin{array}{ll}
2^n\I\sigma_y & \quad\mbox{for odd $n$},\\
2^n\sigma_x & \quad\mbox{for even $n$}.
\end{array}
\right.
\end{equation}
We substitute Eq.~(\ref{hatano-eq4070}) for each even-order term of the right-hand side of Eq.~(\ref{hatano-eq4040}) and arrive at
\begin{equation}\label{hatano-eq4080}
C_1(x)=\frac{1}{2}\sum_{n=0}^\infty a_n(2x)^n\sigma_x
=x\frac{1+\E^{-2x}}{1-\E^{-2x}}\sigma_x
=\left(x\coth x\right) \sigma_x.
\end{equation}
(In the second equality, we used the Taylor expansion~(\ref{hatano-eq4050}) in the reverse direction.)

In summary, we have 
\begin{equation}\label{hatano-eq4090}
\E^{\frac{x}{2}\gamma\sigma_x}\E^{x\sigma_z}
\E^{\frac{x}{2}\gamma\sigma_x}
=\E^{x\left[\sigma_z+\gamma \left(x\coth x\right) \sigma_x+\mathfunc{O}(\gamma^2)\right]}.
\end{equation}
The coefficient $x\coth x$ behaves as shown in Fig.~\ref{hatano-fig40}.
%\begin{figure}
%\end{figure}
We have $x\coth x\simeq 1$ for small $x$ as is expected, but $x\coth x\simeq x$ for large $x$, and hence the first-order perturbation term grows as $x^2$.

\section{Summary}
\label{hatano-sec60}

In the present article, we have reviewed a continual effort on generalization of the Trotter formula to higher-order exponential product formulas.
As was emphasized in Sect.~\ref{hatano-sec20}, the exponential product formula is a good and useful approximant, particularly because it conserves important symmetries of the system dynamics.

We focused on two algorithms of constructing higher-order exponential product formulas.
The first is the fractal decomposition, where we construct higher-order formulas recursively.
The second is to make use of the quantum analysis, where we compute higher-order correction terms directly.
As interludes, we also have described the decomposition of symplectic integrators, the approximation of time-ordered exponentials, and the perturbational composition.
It is our hope that the readers find the present article a useful and tutorial ``manual" when they numerically investigate dynamical systems.
For more practical applications of the exponential product formulas, we refer the readers to the review articles found in Refs.~\cite{hatano-ref410,hatano-ref420,hatano-ref430,hatano-ref440}.

\appendix
\section{Hybrid exponential product formula}
\label{hatano-app05}

We mention here another kind of the exponential product formula~\cite{hatano-ref230,hatano-ref200,hatano-ref250}.\index{hybrid exponential product formula}
Consider the Trotter approximant
\begin{equation}\label{hatano-eq1000}
\E^{xA}\E^{xB}=\E^{x(A+B)+\frac{1}{2}x^2\commut{A}{B}+\mathfunc{O}(x^3)}.
\end{equation}
We can cancel out the second-order correction term in the form
\begin{equation}\label{hatano-eq1010}
\E^{xA}\E^{xB}\E^{-\frac{1}{2}x^2\commut{A}{B}}
=\E^{x(A+B)+\mathfunc{O}(x^3)}.
\end{equation}
If, in some problems, the commutation relation $\commut{A}{B}$ is easily diagonalized, Eq.~(\ref{hatano-eq1010}) may be a useful approximant.

A more complicated one is the fourth-order approximant~\cite{hatano-ref230,hatano-ref200,hatano-ref250}
\begin{equation}\label{hatano-eq1020}
\E^{\frac{x^3}{432}\commut{B}{\commut{A}{B}}}
S_a\left(\frac{x}{3}\right)S_b\left(\frac{x}{3}\right)S_a\left(\frac{x}{3}\right)
\E^{\frac{x^3}{432}\commut{B}{\commut{A}{B}}}
=\E^{x(A+B)+\mathfunc{O}(x^5)},
\end{equation}
where
\begin{equation}\label{hatano-eq1025}
S_a(x)\equiv \E^{\frac{1}{2}xA}\E^{xB}\E^{\frac{1}{2}xA}
\qquad\mbox{and}\qquad
S_b(x)\equiv \E^{\frac{1}{2}xB}\E^{xA}\E^{\frac{1}{2}xB}.
\end{equation}
In fact, the diffusion equation\index{diffusion equation} is described by
\begin{equation}\label{hatano-eq1030}
A=-\frac{1}{2}\Delta
\qquad\mbox{and}\qquad
B=V(\vec{q})
\end{equation}
and we have
\begin{equation}\label{hatano-eq1040}
\commut{B}{\commut{A}{B}}=\left(\nabla V(\vec{q}) \right)^2.
\end{equation}

The above type of the exponential product formula was referred to as the hybrid exponential product formula.
We do not give its details in this article, since commutation relations are not easily diagonalized except for a few specific problems.

\section{World-line quantum Monte Carlo method}
\label{hatano-app10}
\renewcommand{\thesection}{\arabic{chapter}.\arabic{section}}

In the present appendix, we give a short review of the world-line quantum Monte Carlo method.\index{quantum Monte Carlo simulation}\index{world-line quantum Monte Carlo simulation}
The world-line quantum Monte Carlo method is to transform the partition function~(\ref{hatano-eq120}) of a quantum system ${\cal H}$ into the partition function of a classical system by means of the path-integral representation and simulate the latter system.
We explain the method using the transverse Ising model~(\ref{hatano-eq130}), or ${\cal H}=A+B$ with
\begin{equation}\label{hatano-eq2000}
A=-\sum_{\langle i,j \rangle}J_{ij}\sigma^z_i\sigma^z_j
\qquad\mbox{and}\qquad
B=-\Gamma\sum_i\sigma^x_i.
\end{equation}

The starting point is the Trotter decomposition~(\ref{hatano-eq220}) of the partition function, namely the Suzuki-Trotter transformation~\cite{hatano-ref30}, of the form:
\begin{eqnarray}\label{hatano-eq2010}
Z&=&\mathfunc{Tr} \E^{-\beta {\cal H}}=\lim_{n\to\infty}
\mathfunc{Tr}\left(\E^{-\frac{\beta}{n}A}\E^{-\frac{\beta}{n}B}\right)^n
%\nonumber\\
%&=&
=
\lim_{n\to\infty}\sum_{\left\{\sigma_i\right\}}
\left\langle\left\{\sigma_i^{(0)}\right\}\left|
\left(\E^{-\frac{\beta}{n}A}\E^{-\frac{\beta}{n}B}\right)^n
\right|\left\{\sigma_i^{(0)}\right\}\right\rangle
\nonumber\\
&=&
\lim_{n\to\infty}\sum_{\left\{\sigma_i^{(0)}\right\}}
\left\langle\left\{\sigma_i^{(0)}\right\}\right|
\E^{-\frac{\beta}{n}A}\E^{-\frac{\beta}{n}B}
\E^{-\frac{\beta}{n}A}\E^{-\frac{\beta}{n}B}\cdots
%\nonumber\\
%&&
%\phantom{
%\lim_{n\to\infty}\sum%_{\left\{\sigma_i^{(0)}\right\}}
%\left\langle\left\{\sigma_i^{(0)}\right\}\right|
%\E^{-\frac{\beta}{n}A}
%\E^{-\frac{\beta}{n}B}
%}
%\cdots
%\E^{-\frac{\beta}{n}A}
\E^{-\frac{\beta}{n}B}
\left|\left\{\sigma_i^{(0)}\right\}\right\rangle.\qquad\quad
\end{eqnarray}
In the second line, we have taken the trace with respect to a complete basis set by using the spin  $z$ axis as the quantization axis:
\begin{equation}\label{hatano-eq2015}
\sigma^z_k\left|\left\{\sigma_i^{(0)}\right\}\right\rangle=\sigma_k^{(0)}\left|\left\{\sigma_i^{(0)}\right\}\right\rangle,
\end{equation}
where the eigenvalue is $\sigma_k^{(0)}=\pm1$.
The meaning of the superscript $(0)$ becomes self-evident just below.
We further insert the resolution of unity in between each pair of the exponential operators in the last line of Eq.~(\ref{hatano-eq2010}), obtaining
\begin{eqnarray}\label{hatano-eq2020}
\lefteqn{
Z=
\lim_{n\to\infty}\sum_{\left\{\sigma_i^{(0)}\right\}}
\sum_{\left\{\sigma_i^{(1)}\right\}}\sum_{\left\{\sigma_i^{(2)}\right\}}\cdots
\sum_{\left\{\sigma_i^{(n-1)}\right\}}
\left\langle\left\{\sigma_i^{(0)}\right\}\left|
\E^{-\frac{\beta}{n}A}
\right|\left\{\sigma_i^{(0)}\right\}\right\rangle
\left\langle\left\{\sigma_i^{(0)}\right\}\left|
\E^{-\frac{\beta}{n}B}
\right|\left\{\sigma_i^{(1)}\right\}\right\rangle
}
%\nonumber\\
%&&
%\qquad
\nonumber\\
&&
\qquad
\times
\left\langle\left\{\sigma_i^{(1)}\right\}\left|
\E^{-\frac{\beta}{n}A}
\right|\left\{\sigma_i^{(1)}\right\}\right\rangle
\left\langle\left\{\sigma_i^{(1)}\right\}\left|
\E^{-\frac{\beta}{n}B}
\right|\left\{\sigma_i^{(2)}\right\}\right\rangle
%\nonumber\\
%&&
%\qquad
\cdots
%\times
\left\langle\left\{\sigma_i^{(n-1)}\right\}\left|
\E^{-\frac{\beta}{n}B}
\right|\left\{\sigma_i^{(0)}\right\}\right\rangle.
\end{eqnarray}
In the above expression, we used the fact that the operator $A$ is diagonal in the representation of $\{\sigma_i^{(m)}\}$ and hence made the complete set on the both sides of each operator $\E^{-\frac{\beta}{n}A}$ identical.
In contrast, the operator $\E^{-\frac{\beta}{n}B}$ has off-diagonal elements.

Let us calculate the matrix elements in Eq.~(\ref{hatano-eq2020}).
The matrix element of the operator $\E^{-\frac{\beta}{n}A}$ is easy:
\begin{equation}\label{hatano-eq2030}
\left\langle\left\{\sigma_i^{(m)}\right\}\left|
\E^{-\frac{\beta}{n}A}
\right|\left\{\sigma_i^{(m)}\right\}\right\rangle
=\exp\left(\frac{\beta}{n}\sum_{\langle i,j \rangle}J_{ij}\sigma_i^{(m)}\sigma_j^{(m)}\right).
\end{equation}
This is because the operators $\left\{\sigma^z_i\right\}$ are all diagonal in the present representation as in Eq.~(\ref{hatano-eq2015}).
On the other hand, the operator $\E^{-\frac{\beta}{n}B}$ has off-diagonal elements as well as diagonal ones in the following form:
\begin{equation}\label{hatano-eq2040}
\left\langle\left\{\sigma_i^{(m)}\right\}\left|
\E^{-\frac{\beta}{n}B}
\right|\left\{\sigma_i^{(m+1)}\right\}\right\rangle
=\prod_i
\left\langle\sigma_i^{(m)}\left|
\E^{\frac{\beta\Gamma}{n}\sigma^x_i}
\right|\sigma_i^{(m+1)}\right\rangle
\end{equation}
with each matrix element given by
\begin{equation}\label{hatano-eq2050}
%\begin{eqnarray}\label{hatano-eq2050}
%\lefteqn{
\left.\begin{array}{r}
\\
\left\langle\sigma_i^{(m)}\left|
\E^{\frac{\beta\Gamma}{n}\sigma^x_i}
\right|\sigma_i^{(m+1)}\right\rangle
%}
%\nonumber\\
%&=&
=
\end{array}\right.
\begin{array}{cc}
&
\begin{array}{cc}
\left|\sigma_i^{(m+1)}=+1\right\rangle\quad &
\quad\left|\sigma_i^{(m+1)}=-1\right\rangle
\end{array}
\\
\begin{array}{c}
\left\langle\sigma_i^{(m)}=+1\right| \\
\\
\left\langle\sigma_i^{(m)}=-1\right|
\end{array}
&
\left(\begin{array}{cc}
{\displaystyle \qquad\cosh\frac{\beta\Gamma}{n}\qquad} &
{\displaystyle \qquad\sinh\frac{\beta\Gamma}{n}\qquad} \\
&\\
{\displaystyle \qquad\sinh\frac{\beta\Gamma}{n}\qquad} &
{\displaystyle \qquad\cosh\frac{\beta\Gamma}{n}\qquad}
\end{array}\right)
\end{array}.
%\end{eqnarray}
\end{equation}
These matrix elements are expressed in a single equation
\begin{equation}\label{hatano-eq2060}
\left\langle\sigma_i^{(m)}\left|
\E^{\frac{\beta\Gamma}{n}\sigma^x_i}
\right|\sigma_i^{(m+1)}\right\rangle
=\exp\left(
\gamma_n\sigma_i^{(m)}\sigma_i^{(m+1)}+\delta_n
\right),
\end{equation}
where the parameters $\gamma_n$ and $\delta_n$ are defined in
\begin{equation}\label{hatano-eq2070}
\E^{\gamma_n+\delta_n}=\cosh\frac{\beta\Gamma}{n}
\qquad\mbox{and}\qquad
\E^{-\gamma_n+\delta_n}=\sinh\frac{\beta\Gamma}{n},
\end{equation}
or more explicitly defined by
\begin{equation}\label{hatano-eq2080}
\gamma_n=-\frac{1}{2}\log\tanh\frac{\beta\Gamma}{n}
\qquad\mbox{and}\qquad
\delta_n=\frac{1}{2}\log\frac{1}{2}\sinh\frac{2\beta\Gamma}{n}.
\end{equation}
\begin{figure}
\begin{center}
\begin{minipage}[t]{0.35\textwidth}
\vspace{0mm}
\caption{The three-dimensional classical system~(\ref{hatano-eq2100}) mapped from the two-dimensional quantum system~(\ref{hatano-eq2000}).}
\label{hatano-figA10}
\end{minipage}
\hspace{0.05\textwidth}
\begin{minipage}[t]{0.45\textwidth}
\vspace{0mm}
\includegraphics[width=\textwidth]{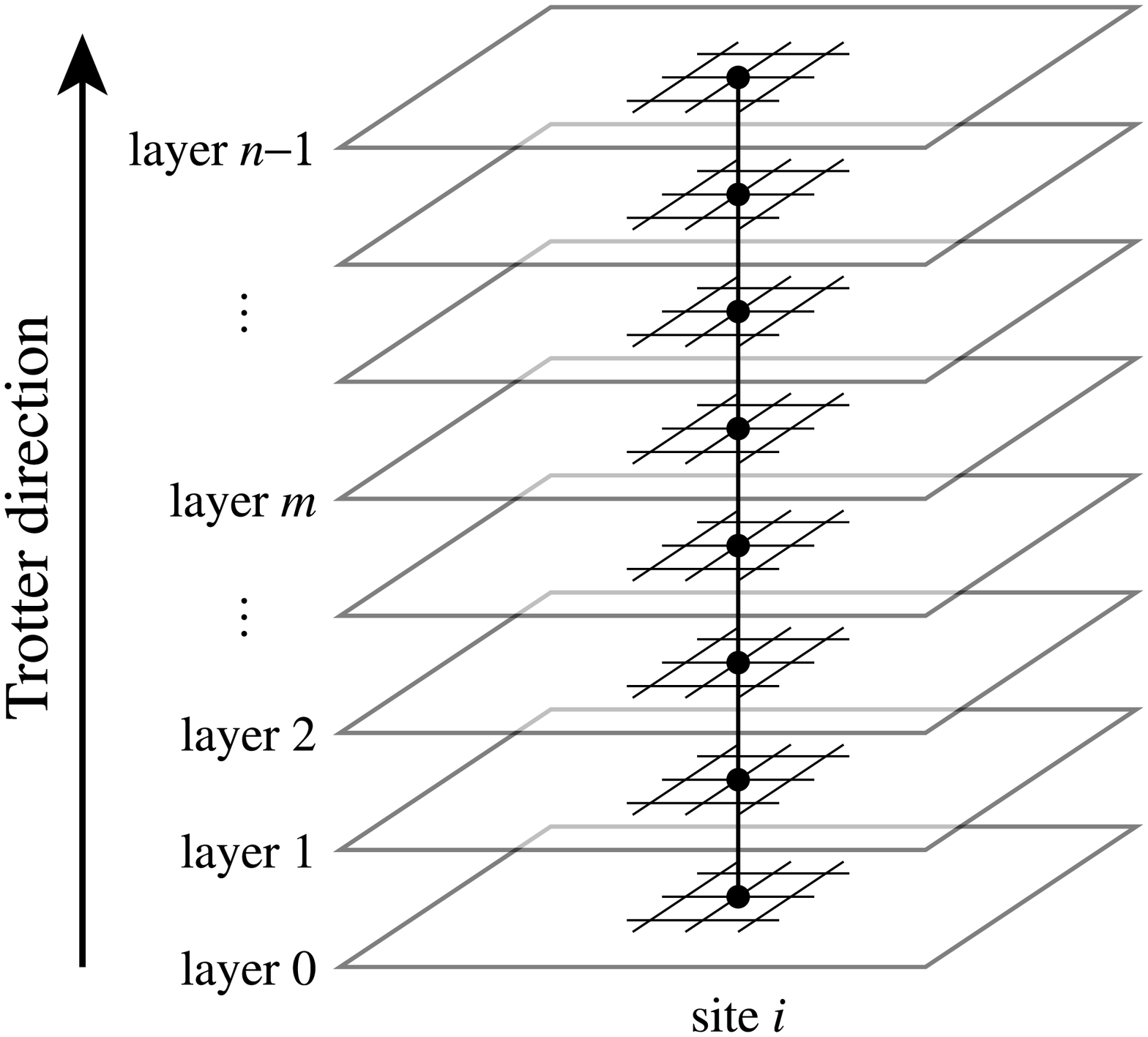}
\end{minipage}
\end{center}
\end{figure}

The expressions~(\ref{hatano-eq2030}) and~(\ref{hatano-eq2060}) give the partition function~(\ref{hatano-eq2020}) in the form~\cite{hatano-ref30}
\begin{equation}\label{hatano-eq2090}
Z=\lim_{n\to\infty}\sum_{\left\{\sigma_i^{(m)}\right\}}\E^{-\beta{\cal H}_n}
\end{equation}
with the resulting classical Hamiltonian~\cite{hatano-ref30}
\begin{equation}\label{hatano-eq2100}
-\beta{\cal H}_n
\equiv
\frac{\beta}{n}\sum_{m=0}^{n-1}\sum_{\langle i,j \rangle}J_{ij}\sigma_i^{(m)}\sigma_j^{(m)}
+\gamma_n\sum_{m=0}^{n-1}\sum_i\sigma_i^{(m)}\sigma_i^{(m+1)},
\end{equation}
where we dropped a constant term due to $\delta_n$.
Note that the periodic boundary conditions, $\sigma_i^{(n)}\equiv\sigma_i^{(0)}$, must be required in the second term of Eq.~(\ref{hatano-eq2100}) because the trace operation in Eq.~(\ref{hatano-eq2010}) demands it.

The classical Hamiltonian~(\ref{hatano-eq2100}) is interpreted as follows (Fig.~\ref{hatano-figA10}).
Suppose that the original quantum system~(\ref{hatano-eq2000}) is defined on a square lattice.
The first term of Eq.~(\ref{hatano-eq2100}) indicates that the two-dimensional system is replicated into $n$ layers with the intra-layer interaction reduced by $n$ times.
The second term of Eq.~(\ref{hatano-eq2100}) represents the inter-layer interactions.
The coupling is $-\gamma_n/\beta$ as defined in Eq.~(\ref{hatano-eq2100}).
Thus the quantum system on a square lattice is mapped to an Ising model on a cubic lattice.
In general, a $d$-dimensional quantum system is mapped to a $(d+1)$-dimensional classical system.
The additional axis is called the Trotter direction.
The physical quantities of the quantum system can be estimated by Monte Carlo simulation of the mapped classical system.
This is the basic idea of the world-line quantum Monte Carlo method~\cite{hatano-ref30}.
\begin{figure}
\begin{center}
\begin{minipage}[t]{0.47\textwidth}
\vspace{0mm}
\includegraphics[width=\textwidth]{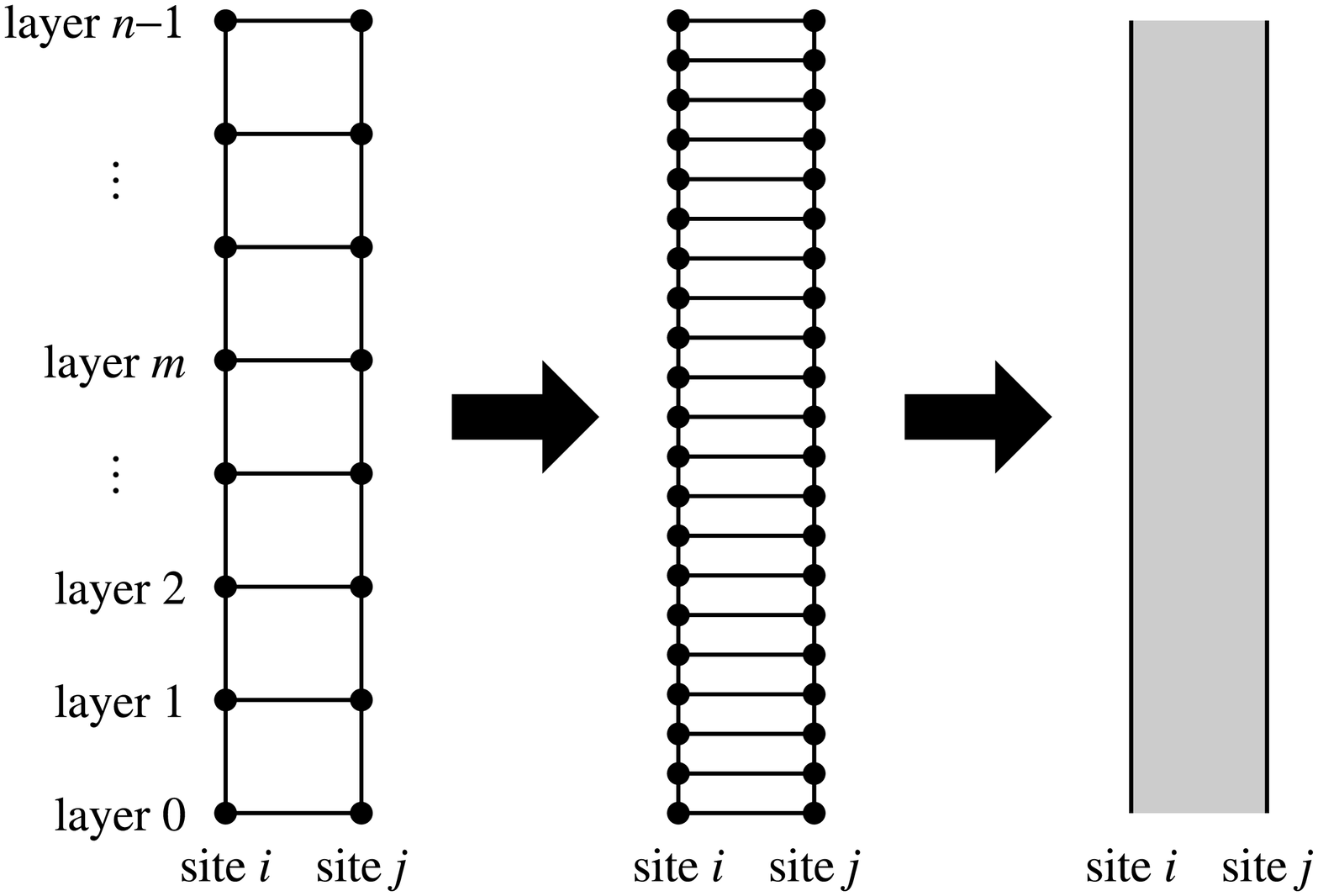}
\caption{In the Trotter limit $n\to\infty$, the Trotter axis becomes a continuum.
The intra-layer interaction becomes an interaction between two continuum axes.}
\label{hatano-figA20}
\end{minipage}
\hspace{0.1\textwidth}
\begin{minipage}[t]{0.33\textwidth}
\vspace{0mm}
\includegraphics[width=\textwidth]{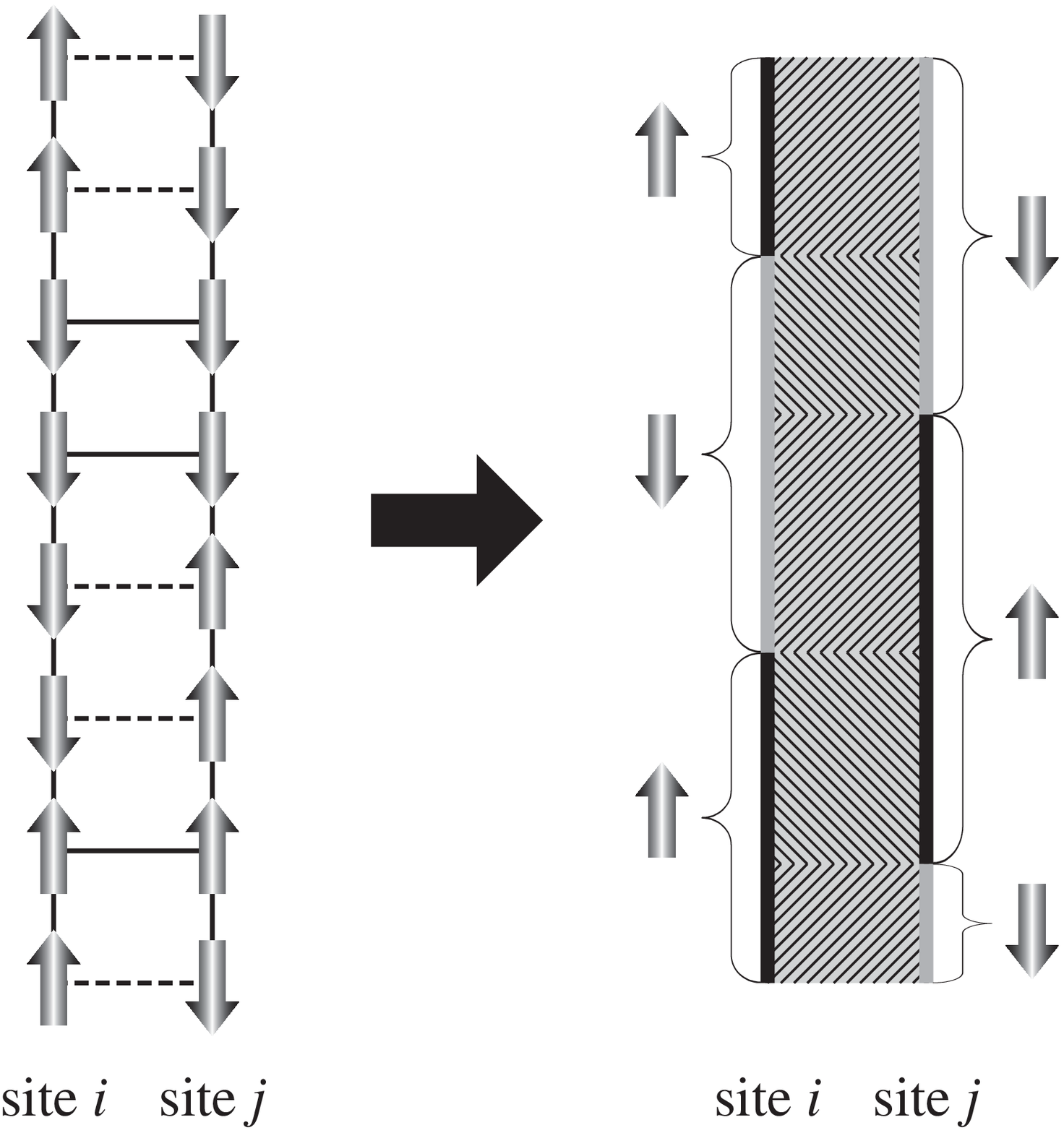}
\caption{Spins on lattice points become domains on Trotter axes in the Trotter limit $n\to\infty$.}
\label{hatano-figA30}
\end{minipage}
\end{center}
\end{figure}

We can use this mapping in order to study the quantum annealing~\cite{hatano-ref380,hatano-ref390,hatano-ref400}.\index{quantum annealing}
Suppose that we look for the ground state of the diagonal part $A$ of the system~(\ref{hatano-eq2000}).
Random exchange interactions $\{J_{ij}\}$ may produce many local minima that are only slightly above the ground state in energy but far apart from the ground state in the phase space.
The simulated annealing, a well-known method of ground-state search, is often trapped in a local minima and does not reach the ground state.
In quantum annealing, we use the transverse field $\Gamma$ in order to induce tunneling from local minima to the ground state.
We first apply the off-diagonal part $B$ of Eq.~(\ref{hatano-eq2000}) strongly and turn it off gradually, hoping to end up with the ground state of the diagonal part $A$.
This corresponds to a Monte Carlo simulation of the mapped classical system~(\ref{hatano-eq2100}) with the intra-layer coupling $\gamma_n$ being infinitesimally weak at the beginning and infinitely strong at the end.
Each layer of the system~(\ref{hatano-eq2100}) is first independent of each other and is gradually frozen into an identical configuration, which we hope is the ground state.

An annoying problem inherent in the algorithm of the quantum Monte Carlo method is the systematic error due to the finite Trotter number $n$.\index{Trotter number}
It used to be that simulations were carried out for various finite values of $n$, quantities were estimated in each simulation, and then the limit $n\to\infty$ was taken in the process of the data analysis, which was called the Trotter extrapolation.\index{Trotter extrapolation}
A recent development of the quantum Monte Carlo method dramatically changed the situation.
We here mention the development briefly;
see Ref.~\cite{hatano-ref370} for a tutorial and exhaustive review of the topic.

In the most recent quantum Monte Carlo algorithm, it is possible for some systems to take the Trotter limit \textit{before} we set up the classical system for simulation.
Taking the Trotter limit $n\to\infty$, we have a continuum Trotter axis (Fig.~\ref{hatano-figA20}).
(Note again that the boundary conditions are required in the Trotter direction.)
The interaction is described as follows (Fig.~\ref{hatano-figA30}).
%\begin{figure}
%\end{figure}
Instead of Ising spins on lattice points of a Trotter axis, we have up-spin domains and down-spin domains on the axis.
Instead of intra-layer interactions between a pair of nearest-neighbor spins, we have parallel-spin areas and anti-parallel-spin areas.
In Monte Carlo simulation, we update the up-spin domains and down-spin domains on the basis of the energy of the parallel-spin areas and anti-parallel-spin areas.

It is thus possible in such situations to carry out a simulation in the Trotter limit $n\to\infty$.
Monte Carlo estimates of such a simulation are free of the systematic error of the order $\beta^2/n$ in Eq.~(\ref{hatano-eq220}), and hence do not need the higher-order exponential product formula for such systems.

\end{document}